\documentclass[useAMS,usenatbib]{mn2e}
\usepackage{color}
\usepackage{graphicx}
\usepackage{lscape}
\usepackage{amssymb,amsmath}
\usepackage{pifont}
\usepackage{url}
\usepackage{float}

\title[Multi-band PL relations for Cepheids in the SMC and LMC]
{Period-Luminosity relations derived from the OGLE-III First-overtone mode Cepheids in the Magellanic Clouds}
\author[Bhardwaj et al.]{Anupam Bhardwaj$^1$\thanks{E-mail:
anupam.bhardwajj@gmail.com}, Chow-Choong Ngeow$^2$, Shashi M. Kanbur$^3$ and
\newauthor  
Harinder P. Singh$^1$
\\
\\
1. Department of Physics \& Astrophysics, University of Delhi, Delhi 110007, India \\
2. Graduate Institute of Astronomy, National Central University, Jhongli 32001, Taiwan \\
3. State University of New York, Oswego, NY 13126, USA\\
}
\begin{document}

\date{Accepted 2016 March 8. Received 2016 March 8; in original form 2015 December 8}

\pagerange{\pageref{firstpage}--\pageref{lastpage}} \pubyear{2015}

\maketitle

\label{firstpage}

\begin{abstract}
We present multi-band Period-Luminosity (PL) relations for first-overtone mode Cepheids in the Small Magellanic Cloud (SMC). We derive optical band PL relations and the Wesenheit function using $VI$ mean magnitudes from the Optical Gravitational Lensing Experiment (OGLE-III) survey. We cross-match OGLE-III first-overtone mode Cepheids to the 2MASS and SAGE-SMC catalogs to derive PL relations at near-infrared ($JHK_s$) and mid-infrared ($3.6~\&~4.5\mu\mathrm{m}$) wavelengths. We test for possible non-linearities in these PL relations using robust statistical tests and find a significant break only in the optical-band PL relations at 2.5 days for first-overtone mode Cepheids. We do not find statistical evidence for a non-linearity in these PL relations at 1 day. The multi-band PL relations for fundamental-mode Cepheids in the SMC also exhibit a break at 2.5 days. We suggest that the period break around 2.5 days is related to sharp changes in the light curve parameters for SMC Cepheids. We also derive new optical and mid-infrared band PL relations for first-overtone mode Cepheids in the Large Magellanic Cloud (LMC). We compare multi-band PL relations for first-overtone mode Cepheids in the Magellanic Clouds and find a significant difference in the slope of the $V$-band PL relations but not for $I$-band PL relations. The slope of PL relations are found to be consistent in most of the infrared bands. A relative distance modulus of $\Delta\mu=0.49\pm0.02$~mag between the two clouds is estimated using multi-band PL relations for the first-overtone mode Cepheids in the SMC and LMC.

\end{abstract}
\section{Introduction}

The Period-Luminosity (PL) relation for Cepheid variables is a vital tool in the cosmic 
distance ladder to obtain distances to the Local Group galaxies and determine an accurate 
and precise value of the Hubble constant \citep{riess09, riess11} that is independent of 
the Cosmic Microwave Background \citep{planck14}. The PL relation or the 
Leavitt law was first introduced by \citet{leavitt12} for Cepheids in the Small Magellanic 
Cloud (SMC). Following this, many studies were published concerning PL relations at 
optical and near-infrared wavelengths for Cepheids in the SMC \citep[for example,]
[and references therein]{welch84, laney86, welch87, cald91,laney94}. 
At optical wavelengths, the second phase of the Optical Gravitational Lensing Experiment 
(OGLE-II) survey derived PL relations for Cepheid variables \citep{udal99}. Later, this catalog 
was used in many studies on PL and/or Period-Color (PC) relations 
\citep{udal99a, sharp02, storm04, tammann09}. Similarly, the PL relations 
for $\sim600$ Cepheids in the SMC were derived by the EROS 
(Exp\'{e}rience de Recherche d'Objets Sombres) collaboration \citep{bauer99}.
Further, \citet{gmat00} used Cepheids in the Magellanic Clouds from OGLE-II and their 
counterparts in 2MASS and DENIS catalogs to derive infrared band PL relations.

In the past decade, a catalog of a large number of Cepheids in the SMC was released by the 
third phase of OGLE survey \citep[OGLE III,][]{oglesmcceph}. The authors provided optical band 
PL and Period-Wesenheit (PW) relations for fundamental and first-overtone mode 
Cepheids but these relations did not take account of extinction corrections. At infrared 
wavelengths, \citet{ngeow10} and \citet{ngeow12} derived PL relations for Cepheids using 
{\it Spitzer} and {\it AKARI} archival data. \citet{tammann09} derived fundamental mode
PL and PC relations for Cepheids in the SMC and \citet{tammann11} extended this work to
first-overtone mode Cepheids and compared these relations with those of Cepheids in metal-poor 
Local Group galaxies. Further, \citet{matsunaga11} used OGLE-III counterparts 
of SMC Cepheids in the IRSF observations to derive PL relations and \citet{inno13} 
used these random phase observations to derive PW relations. Recently, \citet{subra15} 
also derived the PL and PC relations for the OGLE-III Classical Cepheids and investigated the 
structure of the SMC. 

Many studies in the literature concentrate on fundamental mode Cepheid PL relations for their application 
to the distance scale. Therefore, very few of these studies have derived PL relations for the first-overtone 
mode Cepheids in the Magellanic Clouds, specially at infrared wavelengths. Recently, \citet{ngeow15} derived 
multi-band PL relations for fundamental mode Cepheids in the SMC. They also used various test statistics to find 
evidence of a possible non-linearity at 10 days and provided evidence that the SMC PL relations 
are linear. However, they only concentrated on the Cepheid samples with $P>2.5$~days. 
The reason is that the EROS collaboration found a break in the PL relation for Cepheids with periods shorter 
than 2 days \citep{bauer99}. This was later confirmed in many studies \citep[for example,][]{tammann11, subra15}
but at a slightly different period. Further, \citet{bhardwaj14} have shown that both fundamental 
and the first-overtone mode Cepheids in the SMC exhibit break at 2.5 days in PC and Amplitude-Color (A-C) 
relations at maximum and minimum light in optical bands. For these reasons, we extend previous work to 
study multi-band SMC PL relations for the first-overtone mode Cepheids and short period 
fundamental mode Cepheid PL relations to rigorously test for nonlinearity at various periods. 

This paper is structured as follows. In Section~\ref{sec:data}, we discuss the optical, 
near-infrared and mid-infrared data for SMC Cepheid variables used in our analysis along with the 
extinction corrections. We derive multi-band PL relations for first-overtone mode
Cepheids in the SMC in Section~\ref{sec:plpc}, and compare these with published results.
We also test these relations for non-linearity at various periods. We compare breaks in the 
first-overtone mode PL relations with the short period break in fundamental mode SMC Cepheid PL 
relations (\S4). We also compare the first-overtone multi-band PL relations for Cepheids in the
two Magellanic clouds (\S5). Finally, the results and important conclusions of 
this study are discussed in Section~\ref{sec:discuss}.

\section{Data and Extinction Correction}
\label{sec:data}

The photometric mean magnitudes for Cepheids in the SMC at $V$- and $I$-bands are
taken from OGLE-III \citep{oglesmcceph}. There are 2626 fundamental (FU) mode and 
1644 first-overtone (FO) mode Cepheids in SMC as classified by the OGLE-III survey. We 
also derive the optical Wesenheit $W_{V,I}=I-1.55(V-I)$ magnitudes, where the color 
coefficient is obtained using the extinction law given by \citet{card89}. We cross-matched 
the OGLE-III SMC FO mode Cepheids with the 2MASS point source catalog \citep{cutri03} using a 
search radius of $2''$ and obtain the corresponding random-phase $JHK_S$ magnitudes. 
We also cross-matched OGLE-III Cepheids with publicly released SAGE-SMC data and 
obtain IRAC band photometry upto three epochs. The number of matched sources and 
the corresponding mean separations and standard deviations are summarized in 
Table~\ref{table:irac_delta} for the SAGE-SMC data. We estimated the error weighted mean if 
the magnitudes are available for more than one epoch of observation. We find from 
Table~\ref{table:irac_delta} that the majority ($>95\%$) of matched sources are within $1''$ radius
in OGLE-III and SAGE-SMC catalogs. Therefore, the results of our analysis are not affected by the
choice of a greater search radius.

\begin{table}
\begin{minipage}{1.0\hsize}
\caption{Summary of the matched SAGE-SMC archival data for SMC FO mode Cepheids. \label{table:irac_delta}}
\begin{center}
\begin{tabular}{|l|c|c|c|c|c|}
\hline
\hline
Band  &  epoch& $N_{\mathrm{match}}$ & $<\Delta>^{a}$ & $\sigma^b$ & $^c$(in \%) \\
\hline
\hline
3.6$\mu \mathrm{m}$& 0&          790&      0.312&     0.255&      97.34\\
          & 1&         1566&      0.266&     0.248&      98.66\\
          & 2&         1561&      0.310&     0.263&      98.65\\
4.5$\mu \mathrm{m}$& 0&          796&      0.301&     0.260&      97.11\\
          & 1&         1546&      0.263&     0.244&      98.51\\
          & 2&         1550&      0.309&     0.266&      98.52\\
5.8$\mu \mathrm{m}$& 0&          191&      0.327&     0.359&      93.72\\
          & 1&          294&      0.317&     0.378&      95.92\\
          & 2&          289&      0.350&     0.359&      95.85\\
8.0$\mu \mathrm{m}$& 0&          115&      0.329&     0.382&      92.17\\
          & 1&           93&      0.356&     0.399&      90.32\\
          & 2&           75&      0.447&     0.407&      88.00\\
\hline
\end{tabular}
\end{center}
{\footnotesize {$^a~\Delta$ is the separation, in arcsecond, between the matched 
SAGE-SMC archival sources and the OGLE-III SMC Cepheids.}\\
{$^b$~The standard deviation of the mean.}\\
{$^c$~Fraction of matched SAGE-SMC archival sources within $1''$ radius from 
the OGLE-III SMC Cepheids.}} 
\end{minipage}
\end{table}

\begin{figure}
\begin{center}
\includegraphics[width=0.5\textwidth,keepaspectratio]{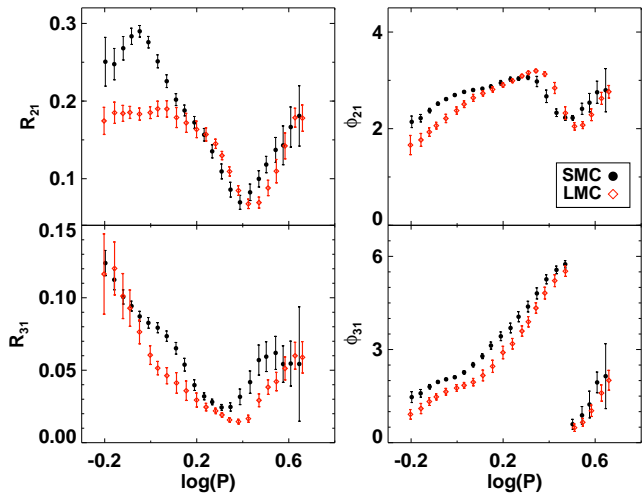}
\caption{A comparison of mean $I$-band Fourier parameters for FO mode Cepheids in the Magellanic Clouds. The error bars
represent $3\sigma$ uncertainties on the mean.} 
\label{fig:fo_fou}
\end{center}
\end{figure}

\begin{figure}
\begin{center}
\includegraphics[width=0.5\textwidth,keepaspectratio]{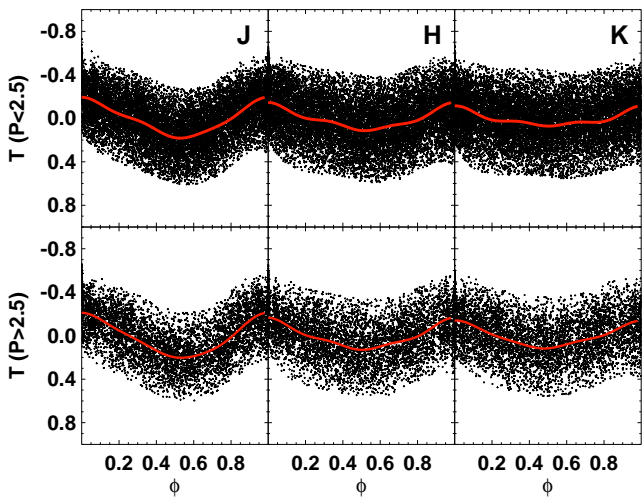}
\caption{Normalised and merged light curves for LMC FO mode Cepheids in period bins P $<$ 2.5 and P $>$ 2.5 days.} 
\label{fig:norm_lc}
\end{center}
\end{figure}
 
We note that near-infrared templates for FO mode Cepheids are not available in literature. 
Recently, \citet{inno15} provided templates for  $J$-band FO mode Cepheids but the calibrator sample
was limited to 10 Cepheids in the SMC. \citet{macri15} derived new near-infrared PL relations for 
FO mode Cepheids in the LMC based on the fairly well sampled light curve data for $\sim500$ Cepheids. 
Therefore, we use time series data from \citet{macri15} together with the methodology presented in 
\cite{sosz05} to correct 2MASS random phase observations to mean magnitudes. We assume negligible
effects of metallicity difference on the light curve shape for FO mode Cepheids in the Magellanic Clouds. It 
is a reasonable assumption considering the fact that FO mode Cepheids have smaller amplitudes and exhibit
near sinusoidal light curve structure. A comparison of $I$-band Fourier parameters for FO mode Cepheids in the
Magellanic Clouds is shown in Fig.~\ref{fig:fo_fou}. We use sliding mean calculations with a step size
of $\log P=0.1$ to estimate mean parameters in each period bin \citep{bhardwaj15}. 
For $\log P>0.2$, the mean amplitude and phase parameters are consistent between the two clouds. There is a 
greater difference in Fourier parameters for $\log P<0.2$ in some period bins but we note that the LMC FO mode 
Cepheid sample is significantly smaller as compared to SMC in this period range. We also emphasize that for 
near-infrared light curves, these differences in amplitude and phase parameters will be even smaller.\\

\begin{figure}
\begin{center}
\includegraphics[width=0.48\textwidth,keepaspectratio]{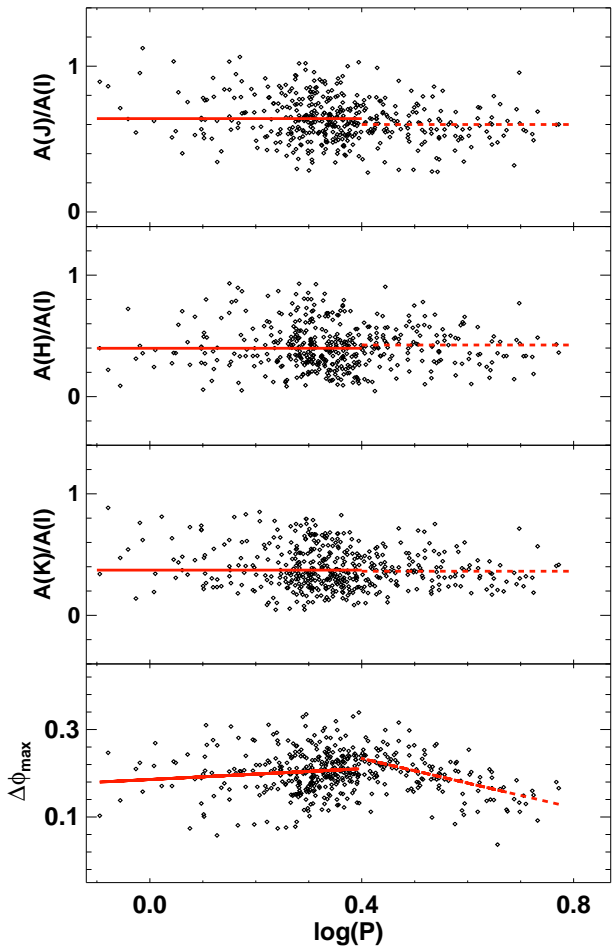}
\caption{Top three panels represent the optical to near-infrared amplitude ratios for LMC FO mode Cepheids and the bottom panel shows 
the phase lag between $I$ and $JHK_s$ light curves for LMC Cepheids.} 
\label{fig:diff}
\end{center}
\end{figure}

We normalize each LMC FO mode Cepheid light curve in such a way 
that the mean magnitude is zero and the amplitude is equal to unity. The Fourier amplitude parameters for 
FO mode Cepheids show a sharp progression around 2.5 days at optical wavelengths. Therefore, we divided our
LMC sample in two period bins and co-added the normalized light curves of each Cepheid in $JHK_s$ bands 
separately. The plot of merged light curves is shown in Fig.~\ref{fig:norm_lc}. Finally, we fit a third-order 
Fourier sine series, $T(\phi) = \sum_{i=1}^{3}A_{i}\sin(2\pi\phi + \Phi_{i})$, to the 
template light curves and obtain amplitude and phase coefficients \citep{bhardwaj15}, listed in Table~\ref{table:fou}. The 
amplitude ratios of near-infrared to optical light curves and the phase lag of maximum light between $I$ and
$JHK_s$ bands for FO mode Cepheids in the LMC are shown in Fig.~\ref{fig:diff}. We do not find any significant 
variation of amplitude ratios as a function of period and adopted median values are listed in Table~\ref{table:amp_i}.
The phase lag for maximum light in $JHK_s$ vs. $I$ for the FO mode Cepheids show a trend as a function of
period for P$>$2.5 days. We note that the $J$ to $V$-band amplitude ratio ($A_J/A_V$) for LMC FO mode Cepheids 
(0.39 for P$<$2.5 and 0.34 for P$>$2.5 days) is similar to the SMC FO mode Cepheids from \citet{inno15}. This is
in contrast to fundamental mode Cepheids where the amplitude ratio for SMC Cepheids are significantly smaller than
LMC Cepheids, and further suggests that the assumption of similar light curve structure for FO Cepheids in the 
Magellanic Clouds is a good approximation.

\begin{table}
\begin{minipage}{1.0\hsize}
\begin{center}
\caption{Fourier coefficients for the near-infrared light curves of FO mode Cepheids. \label{table:fou}}
\begin{tabular}{|c|c|c|c|c|c|c|}
\hline
\hline
	  &  \multicolumn{2}{c}{$J$-band} & \multicolumn{2}{c}{$H$-band} & \multicolumn{2}{c}{$K_s$-band} \\
\hline
{\it i}	&  $A_i$	& $\Phi_i$&	$A_i$	& $\Phi_i$&	$A_i$	& $\Phi_i$\\
\hline
\multicolumn{7}{c}{P $<$ 2.5 days}\\
\hline
      1&      0.171 &     4.538 &     0.108 &     4.586 &     0.071 &     4.625 \\
      2&      0.020 &     5.410 &     0.023 &     5.021 &     0.027 &     4.643 \\
      3&      0.019 &     4.852 &     0.023 &     4.781 &     0.022 &     4.699 \\
\hline
\multicolumn{7}{c}{P $>$ 2.5 days}\\
\hline
      1&      0.197 &     4.530 &     0.128 &     4.661 &     0.115 &     4.728 \\
      2&      0.025 &     5.171 &     0.023 &     4.831 &     0.017 &     4.307 \\
      3&      0.012 &     5.174 &     0.021 &     4.837 &     0.013 &     4.850 \\
\hline
\end{tabular}
\end{center}
\end{minipage}
\end{table}

\begin{table}
\begin{minipage}{1.0\hsize}
\begin{center}
\caption{Near-infrared to optical amplitude ratios for LMC FO mode Cepheids. \label{table:amp_i}}
\begin{tabular}{|c|c|c|c|}
\hline
\hline

P (days) &  $A_J/A_I$	& $A_H/A_I$&	$A_{K_s}/A_I$\\
\hline
	   all&	0.628$\pm$0.007&      0.417$\pm$0.009&      0.398$\pm$0.008\\
	$<2.5$& 0.647$\pm$0.010&      0.413$\pm$0.011&      0.410$\pm$0.011\\
	$>2.5$& 0.594$\pm$0.011&      0.413$\pm$0.013&      0.366$\pm$0.011\\
\hline
\end{tabular}
\end{center}
\end{minipage}
\end{table}

We derive the $I$-band amplitudes for SMC FO mode Cepheids using full phased light curves from OGLE-III.
We use the amplitude ratios for LMC Cepheids together with $I$-band amplitudes for SMC Cepheids and 
estimate the amplitude in near-infrared bands for Cepheids in the SMC.  
Similarly, we calculate the phase of the 2MASS measurement points using the epoch of maximum brightness 
in $I$-band from OGLE-III, i.e. $\phi = \mathrm{frac}\left(\frac{JD^{\lambda} - JD^{I}}{P}\right)$. 
Finally, we calculate the $T(\phi)$ value for this phase and use Fourier coefficients ($A_i$ \& $\Phi_i$) together
with near-infrared amplitudes to estimate the mean magnitudes for FO mode Cepheids using 
$\overline{m(\lambda)} = m(\lambda)_{2MASS} - A(\lambda)\times T(\phi)$.  
We note that the near-infrared magnitudes from 2MASS have large photometric uncertainties which contribute
to the greater dispersion in PL relations. The median uncertainties in the average magnitudes for FO Cepheids
in [$V,\ I,\ J,\ H,\ K_s,\ 3.6\mu\mathrm{m},\ 4.5\mu\mathrm{m}$] bands range from 
$[0.02,\ 0.03,\ 0.06,\ 0.09,\ 0.14,\ 0.06,\ 0.09]$ for P$<$2.5 days to $[0.02,\ 0.03,\ 0.04,\ 0.08,\ 0.08,\ 0.05,\ 0.05]$
for P$>$2.5 days.

The mean magnitudes for Cepheids in multiple bands are corrected for reddening 
using Haschke maps \citep{hasch11}. We provide the input locations for OGLE-III 
Cepheids in terms of RA/DEC and obtain the corresponding E(V-I) color excess.
We correct the magnitudes using the extinction law  $A_{\lambda} = R_{\lambda}E(B-V)$, 
where E(V-I) is related to E(B-V) by the relation E(V-I) = 1.38 E(B-V) \citep{tammann03}. 
For the SMC, the values for total-to-selective absorption ($R_{\lambda}$) are
 $R_{V,\ I,\ J,\ H,\ K,\ 3.6\mu\mathrm{m},\ 4.5\mu\mathrm{m},\ 5.8\mu\mathrm{m},
\ 8.0\mu\mathrm{m}}=\{2.40,\ 1.41,\ 0.69,\ 0.43,\ 0.28,\ 0.12,\ 0.09,\ 0.06,\ 0.04\}$,
corresponding to E(V-I) color excess \citep{ngeow15}.

\section{The Period-Luminosity Relations}
\label{sec:plpc}

\begin{figure}
\begin{center}
\includegraphics[width=0.5\textwidth,keepaspectratio]{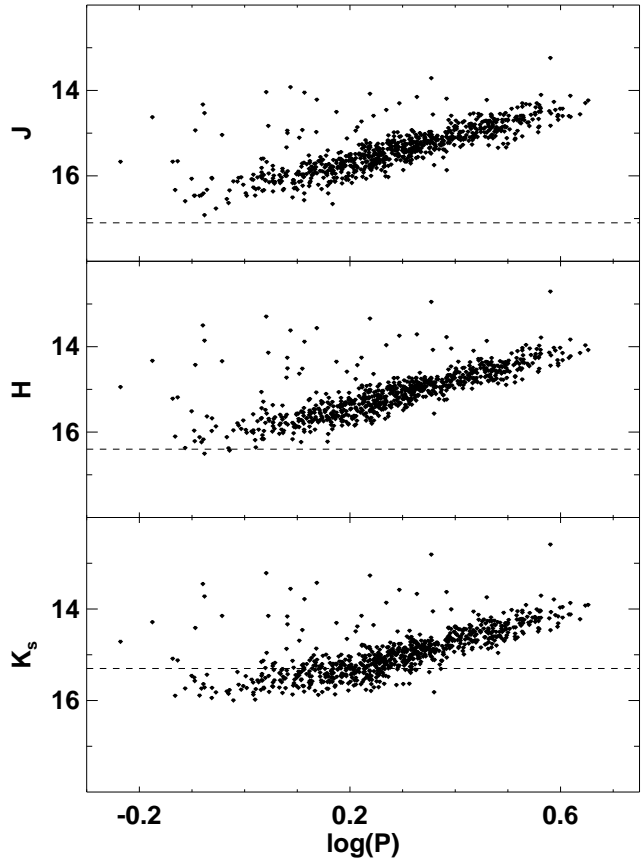}
\caption{Near-infrared band PL relations for SMC FO mode Cepheids. The dashed lines represent the 2MASS $3\sigma$
sensitivity adopted from \citet{cutri03}.} 
\label{fig:fo_pl1}
\end{center}
\end{figure}

\begin{figure}
\begin{center}
\includegraphics[width=0.5\textwidth,keepaspectratio]{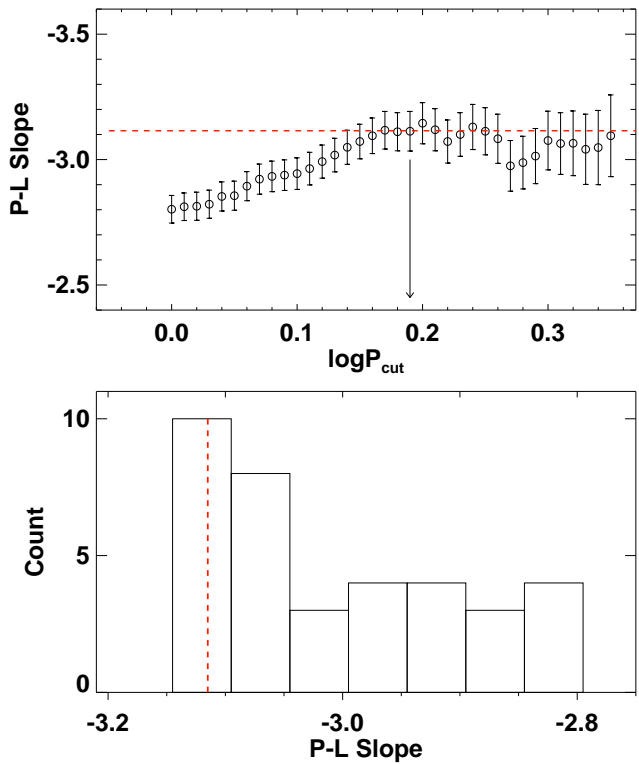}
\caption{The top panel displays slopes of the fitted PL relations for Cepheids having $\log P > \log P_{cut}$, as a 
function of the adopted $\log P_{cut}$. The vertical arrow represents the adopted $\log P_{cut}$ in $K_s$-band.
The bottom panel shows the histogram of the distribution of fitted PL slopes. The (red) 
dashed-lines indicate the mode of the fitted PL slopes based on the histogram. } 
\label{fig:pl_cut}
\end{center}
\end{figure}

\begin{figure}
\begin{center}
\includegraphics[width=0.5\textwidth,keepaspectratio]{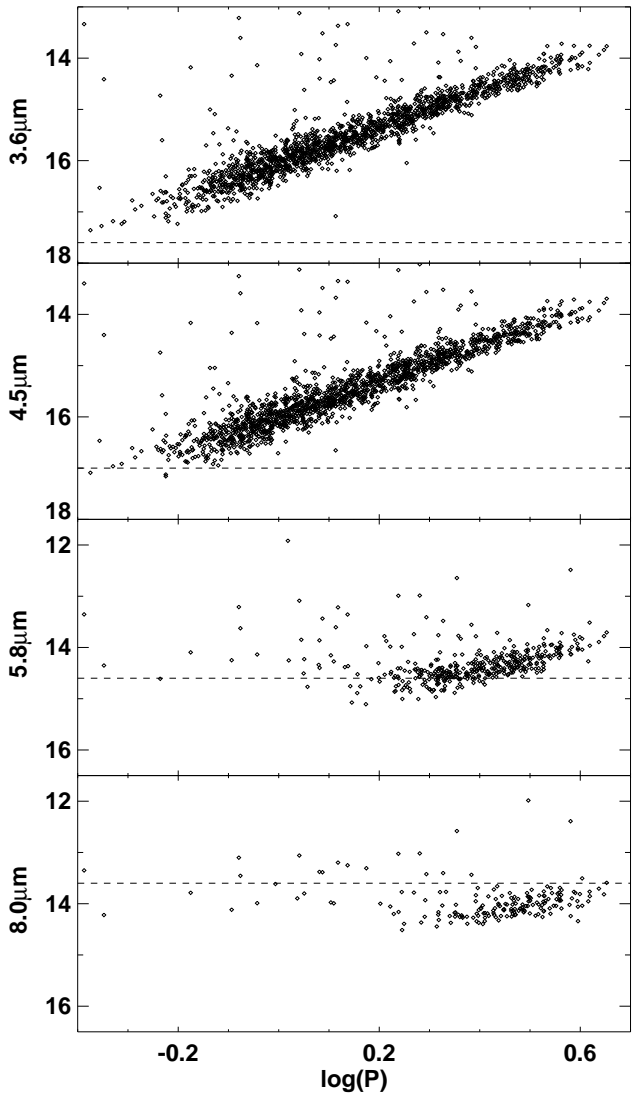}
\caption{Mid-infrared PL relations for SMC FO mode Cepheids. The dashed lines represent the detection limits in each band for SAGE-SMC catalog.}
\label{fig:fo_pl2}
\end{center}
\end{figure}

\citet{oglesmcceph} derived the PL and PW relations for FU and FO mode Cepheids in the SMC at optical wavelengths
in their data release paper. However, these relations were not corrected for extinction. There are many studies 
in the past decade on FU mode Cepheid PL relations (for references, see \S1) but not for FO mode Cepheids in the 
Magellanic Clouds. \citet{gmat00} derived PL relations for FO mode Cepheids by combining OGLE-II data with DENIS and 
2MASS infrared data. \citet{bono02} carried out a study on theoretical and observed PL relations for FO mode
Cepheids in $IK_s$-bands. A detailed study based on PL and PC relations for SMC Cepheids using OGLE catalogs 
was carried out by \citet{tammann09} and  \citet{tammann11}. Recently, \citet{subra15} provided new PL relations for FO 
Cepheids using OGLE-III data and found evidence of several breaks in optical PL and PC relations but no 
statistical tests were done in their analysis. We extend these studies to derive multi-band PL relations and 
use our test statistics \citep{cpapir3}
to rigorously determine any possible non-linearities in these relations for FO mode Cepheids.

Fig.~\ref{fig:fo_pl1} displays near-infrared PL relations using random phase corrected mean magnitudes. 
We find that all $J$ and $H$-band magnitudes are above the detection limit while the $K_s$-band magnitudes 
are influenced due to the incompleteness bias at the short-period end. This incompleteness bias is related to 
well-known Malmquist bias \citep[see,][ for more details]{sandage88}. Therefore, an appropriate period cut is
required in $K_s$-band. We use the method discussed in \citet{ngeow15} to adopt period cuts. In brief, we start
with an initial $\log P_{cut}=0$ and fit a PL relation to the remaining sample with iterative $2.5\sigma$ clipping and
obtain the slope. We repeat this process with a bin size of $\Delta \log P_{cut} = 0.01$ upto a maximum value
of $\log P_{cut}=0.35$.
We plot the slopes of the PL relations as a function of $\log P_{cut}$ in the
upper panel of Fig.~\ref{fig:pl_cut} and present these distributions as histograms in the lower panel.
We determine the mode of the histogram and the absolute difference
between the mode value and the slopes of the PL relations. The $\log P_{cut}$ value corresponding to
the smallest absolute difference is adopted as the final period cut for the $K_s$-band PL relation.
Therefore, we adopt the final period cut at $\log P_{cut}=0.19$ for the $K_s$-band PL relation.
Similarly, Fig.~\ref{fig:fo_pl2} displays the mid-infrared PL relations for FO mode Cepheids. Again, the
$3.6\mu \mathrm{m}$ and $4.5\mu \mathrm{m}$ band PL relations are not influenced by the incompleteness bias near their detection limit.
The magnitudes in $5.8\mu \mathrm{m}$ and $8.0\mu \mathrm{m}$ are mostly below the detection limit and therefore, we will not
consider these bands further in our analysis. 

\begin{figure}
\begin{center}
\includegraphics[width=0.5\textwidth,keepaspectratio]{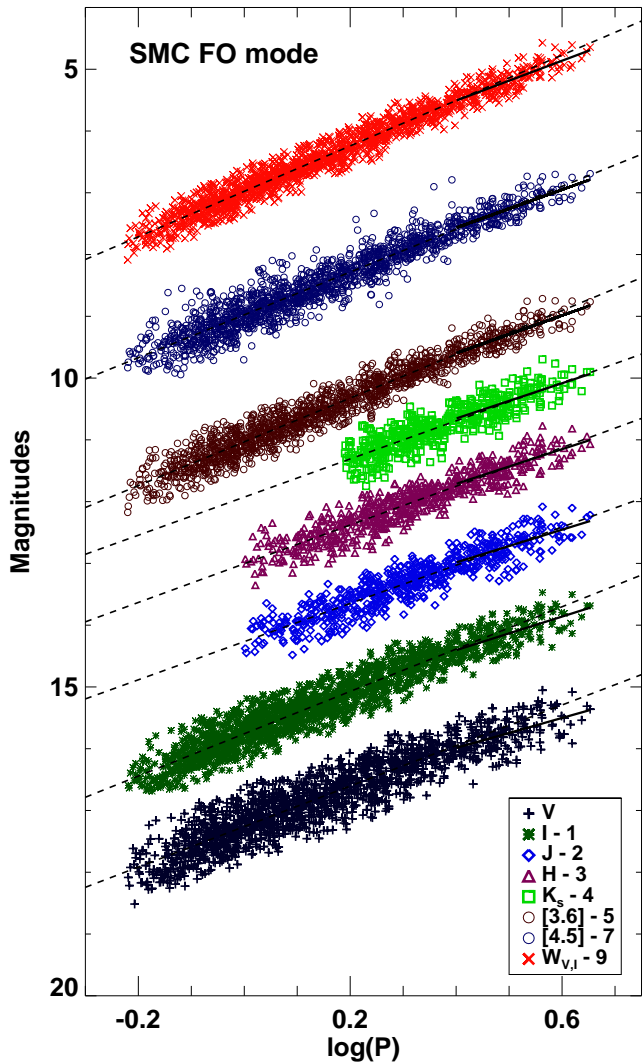}
\caption{Multi-band PL relations and optical Wesenheit relation for SMC FO mode Cepheids.
The solid/dashed lines represent the best fit regression for Cepheids with periods below and above 2.5 days.}
\label{fig:final_pl}
\end{center}
\end{figure}

We make use of extinction corrected mean magnitudes in $VIJHK_s$ \& $3.6/4.5\mu \mathrm{m}$ wavelengths 
to derive PL relations in each band, separately. At optical and mid-infrared wavelengths,
we restrict our sample to $P>0.6$~days because there are very few stars below this period. Similarly,
we only consider stars with $P>1$~day at near-infrared wavelengths. We apply recursive
$2.5\sigma$ clipping to fit a PL relation in each band, where $\sigma$ is the dispersion
in the PL relation after each iteration. Stars that are removed because of this procedure may appear as outliers due to several
reasons including misidentification in the OGLE-III catalog and blending with nearby sources. 
Further investigation implies that the majority of these outliers are present in more than one band and
therefore should be removed before fitting a PL relation. However,
we will not investigate the nature of these outliers in this study. 
Our final PL relations at multiple bands for the SMC FO mode Cepheids are presented in Fig.~\ref{fig:final_pl}
and the results are given in Table~\ref{table:final_pl}.

\begin{table}
\begin{minipage}{1.0\hsize}
\begin{center}
\caption{Multi-band PL relations for SMC FO mode Cepheids. \label{table:final_pl}}
\begin{tabular}{|l|c|c|c|c|}
\hline
\hline
Band	&	PL Slope& 	PL ZP	&	$\sigma$& 	$N$\\
\hline
\hline
                 $V$&     -3.147~$\pm$~0.035     &     17.254~$\pm$~0.008     &      0.268&         1531\\
                 $I$&     -3.332~$\pm$~0.028     &     16.754~$\pm$~0.007     &      0.218&         1531\\
                 $J$&     -3.095~$\pm$~0.055     &     16.267~$\pm$~0.019     &      0.198&          595\\
                 $H$&     -3.151~$\pm$~0.053     &     16.012~$\pm$~0.018     &      0.191&          598\\
             $K_{s}$&     -3.132~$\pm$~0.083     &     15.941~$\pm$~0.032     &      0.197&          457\\
         $3.6\mu \mathrm{m}$&     -3.510~$\pm$~0.024     &     16.030~$\pm$~0.006     &      0.180&         1499\\
         $4.5\mu \mathrm{m}$&     -3.451~$\pm$~0.028     &     15.975~$\pm$~0.007     &      0.200&         1529\\
           $W_{V,I}$&     -3.626~$\pm$~0.020     &     15.974~$\pm$~0.005     &      0.153&         1526\\
\hline
\end{tabular}
\end{center}
\end{minipage}
\end{table}

We note that the dispersion in the FO mode Cepheid PL relations are very similar to FU mode Cepheid PL 
relations \citep{ngeow15}, except at mid-infrared wavelengths. We do not observe any significant
reduction in dispersion of the near-infrared PL relations with or without random phase corrections. For
the $J$-band, the dispersion reduces by $\sim0.01$~mag, while it is negligible in the case of $H$ and $K_s$-band
PL relations. We suspect that the dispersion in the 2MASS PL relations is dominated by photometric errors,
most likely crowding, and the high intrinsic line-of-sight depth of the SMC \citep{subra15}. Hence, phase corrections do not
contribute significantly, especially when the pulsation amplitudes are very small. However, the dispersion in the near-infrared
PL relation for LMC FO mode Cepheids ranges from 0.09-0.14~mag \citep{macri15} as opposed to 0.18-0.20~mag for SMC FO mode Cepheids.
Therefore, we test the contribution of random phase corrections to LMC FO mode Cepheid PL relations. We randomly sample the LMC FO
light curves and estimate the dispersion in each band. We find that the random phase PL relations in $JHK_s$ bands 
do not give significantly different dispersion than the mean light PL relations, specially in $H$ and $K_s$-bands. The
greatest difference of $\sim0.014$~mag occurs for $J$-band in 50 different samples of random PL relations.

\subsection{Comparison with Published PL Relations}

We compare our multi-band PL relations for FO mode Cepheids in the SMC with published results 
\citep{bauer99, gmat00, oglesmcceph, tammann11, inno13, subra15}. The optical band PL relations in our study
are essentially an updated version of PL relations derived by \citet{oglesmcceph} as we account 
for the interstellar extinction. We note that the Wesenheit function, that is independent of the extinction, 
provides precisely the same slope, as expected, in the two set of PL relations. We use a
standard t-test to check the significance of the difference in the slopes of our PL relations with
published work. The details of t-test can be found in \citet{ngeow15} and \citet{cpapir2}. In brief,
we calculate T-values using the errors on the slopes and the standard deviation of PL relations. 
The theoretical T-values are obtained from the t-distribution for an adopted significance level ($\alpha=0.05$),
under the null hypothesis of the two slopes being equal. The probability of acceptance of the
null hypothesis ($p(t)$) of the observed t-statistic ($|T|$) is listed in Table~\ref{table:comp_smc}.
The null hypothesis that the two slopes under consideration are consistent with each other is rejected if $p(t)<\alpha$.

\begin{table}
\begin{minipage}{1.0\hsize}
\begin{center}
\caption{Comparison of SMC FO mode PL relations. \label{table:comp_smc}}
\begin{tabular}{|c|c|c|c|c|c|}
\hline
\hline
	PL Slope& 	$\sigma$& 	$N$&	Src&	$|T|$&	$p(t)$\\
\hline
\hline
\multicolumn{6}{c}{$V$-band}\\
\hline
    -3.147$\pm$0.035     &      0.268&    1531 &TW& ---& ---\\
    -3.060$\pm$0.080     &      0.250&     239 &E99&      0.948&      0.343\\
    -3.171$\pm$0.034     &      0.280&    1644 &S10&      0.491&      0.624\\
    -3.203$\pm$0.060     &      0.250&     597 &T11&      0.779&      0.436\\
    -3.030$\pm$0.056     &      0.270&    1041 &S15&      1.776&      0.076\\
\hline
\multicolumn{6}{c}{$I$-band}\\
\hline
    -3.332$\pm$0.028     &      0.218&    1531 &TW& ---& ---\\
    -3.349$\pm$0.027     &      0.230&    1644 &S10&      0.436&      0.663\\
    -3.374$\pm$0.046     &      0.190&     595 &T11&      0.728&      0.467\\
    -3.230$\pm$0.045     &      0.220&    1041 &S15&      1.930&      0.054\\
\hline
\multicolumn{6}{c}{$J$-band}\\
\hline
    -3.095$\pm$0.055     &      0.198&     595 &TW& ---& ---\\
    -3.104$\pm$0.159     &      0.183&     158 &G00&      0.094&      0.926\\
\hline
\multicolumn{6}{c}{$H$-band}\\
\hline
    -3.151$\pm$0.053     &      0.191&     598 &TW& ---& ---\\
    -3.199$\pm$0.142     &      0.162&     162 &G00&      0.316&      0.754\\
\hline
\multicolumn{6}{c}{$K_s$-band}\\
\hline
    -3.132$\pm$0.083     &      0.197&     457 &TW& ---& ---\\
    -3.102$\pm$0.155     &      0.178&     156 &G00&      0.136&      0.887\\
\hline
\multicolumn{6}{c}{$W_{V,I}$}\\
\hline
    -3.626$\pm$0.020     &      0.153&    1526 &TW& ---& ---\\
    -3.558$\pm$0.025     &      0.126&     688 &G00&      1.972&      0.049\\
    -3.623$\pm$0.020     &      0.160&    1644 &S10&      0.106&      0.915\\
    -3.599$\pm$0.027     &      0.140&    1465 &I13&      0.790&      0.430\\
\hline
\end{tabular}
\end{center}
{\footnotesize \textbf{Notes}: Source : TW - this work; E99 - \citet[EROS collaboration,][]{bauer99}; G00 - \citet{gmat00};
S10 - \citet{oglesmcceph}; T11 - \citet{tammann11}; I13 - \citet{inno13}; S15 - \citet{subra15}.}
\end{minipage}
\end{table}

We find that all PL relations derived in this paper for SMC FO mode Cepheids are consistent with previous studies.
At optical wavelengths, the $V$-band PL relation from \citet{bauer99} is derived in the $V_{\it EROS}$ filter. The PL 
relations taken from \citet{tammann11} are restricted to the short period range ($\log P < 0.4$), whilst those taken
from \citet{subra15} are limited to long period ($\log P > 0.029$) FO mode Cepheids in OGLE-III catalog. Similarly, the PL 
relations adopted from \citet{gmat00}, derived using 2MASS and DENIS data, are limited to long period 
($\log P > 0.3$) Cepheids only. For each set of slopes under consideration, the t-test suggests that the data are 
consistent with the null hypothesis at the $95\%$ significance level. Therefore, the corresponding set of slopes are equal within their 
quoted uncertainties. The slope of the Wesenheit function derived by \citet{gmat00} is marginally inconsistent 
according to the t-test, presumably due to a significantly smaller sample size of OGLE-II Cepheids as compared 
to OGLE-III.

\subsection{A Test for Non-Linearity in PL Relations}

\begin{table*}
\begin{minipage}{1.0\hsize}
\begin{center}
\caption{Results of F and random walk tests for PL, PW and PC relations for FO mode Cepheids in the SMC to test for non-linearity 
at various periods.}
\label{table:fo_smc_frw}
\begin{tabular}{|l|c|c|c|c|c|c|c|c|c|c|c|}
\hline
\hline
$Y$& PL Slope$_{S}$& PL ZP$_{S}$& $\sigma_{S}$& $N_{S}$ & PL Slope$_{L}$& PL ZP$_{L}$& $\sigma_{S}$& $N_{S}$ & $F$ & $p(F)$& $p(R)$\\
\hline
\hline
\multicolumn{12}{c}{For all $P$ with a break at 2.5 days \citep{bhardwaj14}}\\
\hline
                 $V$&    -3.279$\pm$0.046     &    17.260$\pm$0.008     &     0.268&        1349&    -2.426$\pm$0.303     &    16.957$\pm$0.149     &     0.244&         182&    11.104&     0.000&     0.002\\
                 $I$&    -3.427$\pm$0.038     &    16.758$\pm$0.007     &     0.219&        1348&    -2.718$\pm$0.243     &    16.492$\pm$0.119     &     0.197&         183&     9.075&     0.000&     0.010\\
                 $J$&    -3.092$\pm$0.094     &    16.266$\pm$0.024     &     0.197&         423&    -2.649$\pm$0.254     &    16.043$\pm$0.125     &     0.198&         172&     1.599&     0.203&     0.278\\
                 $H$&    -3.136$\pm$0.093     &    16.005$\pm$0.024     &     0.195&         426&    -2.803$\pm$0.231     &    15.833$\pm$0.114     &     0.181&         172&     1.064&     0.346&     0.430\\
             $K_{s}$&    -3.086$\pm$0.212     &    15.927$\pm$0.064     &     0.204&         285&    -2.854$\pm$0.236     &    15.796$\pm$0.116     &     0.184&         172&     0.725&     0.485&     0.391\\
         $3.6\mu \mathrm{m}$&    -3.544$\pm$0.033     &    16.032$\pm$0.006     &     0.185&        1309&    -3.052$\pm$0.181     &    15.817$\pm$0.089     &     0.146&         190&     3.013&     0.049&     0.573\\
         $4.5\mu \mathrm{m}$&    -3.466$\pm$0.038     &    15.976$\pm$0.007     &     0.208&        1339&    -3.085$\pm$0.178     &    15.800$\pm$0.087     &     0.145&         190&     1.091&     0.336&     0.224\\
           $W_{V,I}$&    -3.676$\pm$0.027     &    15.977$\pm$0.005     &     0.154&        1335&    -3.229$\pm$0.175     &    15.799$\pm$0.086     &     0.143&         191&     5.642&     0.004&     0.016\\
               $V-I$&     0.157$\pm$0.011     &     0.503$\pm$0.002     &     0.064&        1339&     0.377$\pm$0.073     &     0.426$\pm$0.036     &     0.058&         181&    13.915&     0.000&     0.000\\
\hline
\multicolumn{12}{c}{For all $P$ with a break at 1 day \citep{subra15}}\\
\hline
                 $V$&    -3.868$\pm$0.246     &    17.218$\pm$0.023     &     0.277&         419&    -3.015$\pm$0.049     &    17.216$\pm$0.014     &     0.262&        1112&     9.956&     0.000&     0.001\\
                 $I$&    -3.677$\pm$0.205     &    16.743$\pm$0.019     &     0.229&         419&    -3.241$\pm$0.040     &    16.727$\pm$0.011     &     0.212&        1112&     5.666&     0.004&     0.011\\
         $3.6\mu \mathrm{m}$&    -3.510$\pm$0.191     &    16.035$\pm$0.018     &     0.210&         404&    -3.487$\pm$0.032     &    16.023$\pm$0.009     &     0.169&        1095&     0.451&     0.637&     0.576\\
         $4.5\mu \mathrm{m}$&    -3.129$\pm$0.220     &    15.998$\pm$0.021     &     0.237&         413&    -3.473$\pm$0.038     &    15.981$\pm$0.010     &     0.186&        1116&     1.573&     0.208&     0.221\\
           $W_{V,I}$&    -3.607$\pm$0.150     &    15.988$\pm$0.014     &     0.170&         415&    -3.567$\pm$0.028     &    15.956$\pm$0.008     &     0.147&        1111&     3.969&     0.019&     0.014\\
               $V-I$&    -0.048$\pm$0.061     &     0.485$\pm$0.006     &     0.069&         418&     0.219$\pm$0.012     &     0.494$\pm$0.003     &     0.060&        1102&    12.427&     0.000&     0.000\\
\hline
\multicolumn{12}{c}{For all $P$ with a break at 1.6 days, related to a feature in Fourier parameters \citep[see, Fig. 8 in][]{bhardwaj14}}\\
\hline
                 $V$&    -3.342$\pm$0.084     &    17.259$\pm$0.009     &     0.273&         956&    -2.805$\pm$0.098     &    17.132$\pm$0.036     &     0.253&         575&     8.472&     0.000&     0.001\\
                 $I$&    -3.443$\pm$0.069     &    16.758$\pm$0.007     &     0.224&         958&    -3.058$\pm$0.078     &    16.653$\pm$0.029     &     0.204&         573&     6.902&     0.001&     0.011\\
                 $J$&    -2.497$\pm$0.314     &    16.202$\pm$0.042     &     0.207&         146&    -3.053$\pm$0.084     &    16.247$\pm$0.032     &     0.194&         449&     2.335&     0.098&     0.269\\
                 $H$&    -2.668$\pm$0.319     &    15.951$\pm$0.043     &     0.211&         147&    -3.155$\pm$0.079     &    16.008$\pm$0.031     &     0.184&         451&     1.386&     0.251&     0.421\\
             $K_{s}$&   -11.005$\pm$12.711    &    17.365$\pm$2.482     &     0.138&          10&    -3.163$\pm$0.085     &    15.951$\pm$0.033     &     0.197&         447&     1.790&     0.168&     0.397\\
         $3.6\mu \mathrm{m}$&    -3.498$\pm$0.061     &    16.033$\pm$0.006     &     0.194&         928&    -3.389$\pm$0.061     &    15.982$\pm$0.023     &     0.157&         571&     1.842&     0.159&     0.573\\
         $4.5\mu \mathrm{m}$&    -3.316$\pm$0.071     &    15.978$\pm$0.007     &     0.220&         947&    -3.368$\pm$0.071     &    15.936$\pm$0.026     &     0.165&         582&     3.581&     0.028&     0.237\\
           $W_{V,I}$&    -3.678$\pm$0.049     &    15.977$\pm$0.005     &     0.160&         948&    -3.452$\pm$0.055     &    15.909$\pm$0.020     &     0.141&         578&     5.184&     0.006&     0.015\\
               $V-I$&     0.139$\pm$0.020     &     0.502$\pm$0.002     &     0.066&         952&     0.279$\pm$0.023     &     0.470$\pm$0.008     &     0.058&         568&    10.083&     0.000&     0.000\\
\hline
\multicolumn{12}{c}{For $P>1$~day with a break at 2.5 days}\\
\hline
                 $V$&    -3.178$\pm$0.075     &    17.240$\pm$0.016     &     0.263&         930&    -2.426$\pm$0.303     &    16.957$\pm$0.149     &     0.244&         182&     5.415&     0.005&     0.054\\
                 $I$&    -3.364$\pm$0.061     &    16.745$\pm$0.013     &     0.214&         929&    -2.718$\pm$0.243     &    16.492$\pm$0.119     &     0.197&         183&     5.082&     0.006&     0.056\\
                 $J$&    -3.092$\pm$0.094     &    16.266$\pm$0.024     &     0.197&         423&    -2.649$\pm$0.254     &    16.043$\pm$0.125     &     0.198&         172&     1.599&     0.203&     0.271\\
                 $H$&    -3.136$\pm$0.093     &    16.005$\pm$0.024     &     0.195&         426&    -2.803$\pm$0.231     &    15.833$\pm$0.114     &     0.181&         172&     1.064&     0.346&     0.432\\
             $K_{s}$&    -3.086$\pm$0.212     &    15.927$\pm$0.064     &     0.204&         285&    -2.854$\pm$0.236     &    15.796$\pm$0.116     &     0.184&         172&     0.725&     0.485&     0.387\\
         $3.6\mu \mathrm{m}$&    -3.542$\pm$0.051     &    16.032$\pm$0.011     &     0.174&         904&    -3.052$\pm$0.181     &    15.817$\pm$0.089     &     0.146&         190&     2.926&     0.054&     0.166\\
         $4.5\mu \mathrm{m}$&    -3.538$\pm$0.060     &    15.991$\pm$0.013     &     0.194&         925&    -3.085$\pm$0.178     &    15.800$\pm$0.087     &     0.145&         190&     2.092&     0.124&     0.151\\
           $W_{V,I}$&    -3.628$\pm$0.043     &    15.965$\pm$0.009     &     0.147&         920&    -3.229$\pm$0.175     &    15.799$\pm$0.086     &     0.143&         191&     3.119&     0.045&     0.062\\
               $V-I$&     0.168$\pm$0.017     &     0.501$\pm$0.004     &     0.060&         921&     0.377$\pm$0.073     &     0.426$\pm$0.036     &     0.058&         181&     9.050&     0.000&     0.001\\
\hline
\multicolumn{12}{c}{For $P<2.5$~days with a break at 1 day}\\
\hline
                 $V$&    -3.868$\pm$0.246     &    17.218$\pm$0.023     &     0.277&         419&    -3.178$\pm$0.075     &    17.240$\pm$0.016     &     0.263&         930&     4.023&     0.018&     0.127\\
                 $I$&    -3.677$\pm$0.205     &    16.743$\pm$0.019     &     0.229&         419&    -3.364$\pm$0.061     &    16.745$\pm$0.013     &     0.214&         929&     1.434&     0.239&     0.568\\
         $3.6\mu \mathrm{m}$&    -3.512$\pm$0.190     &    16.035$\pm$0.018     &     0.210&         405&    -3.542$\pm$0.051     &    16.032$\pm$0.011     &     0.174&         904&     0.020&     0.980&     0.701\\
         $4.5\mu \mathrm{m}$&    -3.135$\pm$0.220     &    15.998$\pm$0.021     &     0.237&         414&    -3.538$\pm$0.060     &    15.991$\pm$0.013     &     0.194&         925&     2.146&     0.117&     0.128\\
           $W_{V,I}$&    -3.607$\pm$0.150     &    15.988$\pm$0.014     &     0.170&         415&    -3.628$\pm$0.043     &    15.965$\pm$0.009     &     0.147&         920&     1.192&     0.304&     0.609\\
               $V-I$&    -0.048$\pm$0.061     &     0.485$\pm$0.006     &     0.069&         418&     0.168$\pm$0.017     &     0.501$\pm$0.004     &     0.060&         921&     6.797&     0.001&     0.011\\
\hline
\end{tabular}
\end{center}
{\footnotesize \textbf{Notes}: 
The subscripts $S$ and $L$ stand for short and long period range, respectively, for a given break period. ZP and $\sigma$ 
represents the zero point and dispersion of the PL relation, respectively. $N$ is the number of Cepheids used in 
deriving the PL relations. $p(F)$ and $p(R)$ represent the probability of acceptance of the null hypothesis i.e. linear relation.}
\end{minipage}
\end{table*}
\begin{table*}
\begin{minipage}{1.0\hsize}
\begin{center}
\caption{Results of the testimator on PL, PW and PC relations for FO mode Cepheids in the SMC.}
\label{table:fo_smc_tm}
\begin{tabular}{|l|c|c|c|c|c|c|c|c|c|c|}
\hline
\hline
Band &$n$  &  $\log(P)$&	$N$&	$\hat{\beta}$&	$\beta_{0}$&	$|t_{obs}|$&	$t_{c}$&	$k$&	Decision&	$\beta_{w}$\\
\hline
\hline
$V$&      1&  -0.21900$-$-0.07000  &         208&    -4.179~$\pm$~0.491     &       ---&       ---&       ---&       ---&       ---         &       ---\\
& 2&  -0.07000$-$0.00100   &211&    -3.069~$\pm$~0.917     &    -4.179&     1.210&     2.717&     0.446&Accept~$H_{0}$&    -3.685\\
& 3&   0.00100$-$0.07100   &210&    -2.834~$\pm$~0.964     &    -3.685&     0.883&     2.717&     0.325&Accept~$H_{0}$&    -3.408\\
& 4&   0.07100$-$0.15000   &209&    -3.144~$\pm$~0.739     &    -3.408&     0.358&     2.717&     0.132&Accept~$H_{0}$&    -3.373\\
& 5&   0.15000$-$0.24600   &211&    -3.637~$\pm$~0.656     &    -3.373&     0.403&     2.717&     0.148&Accept~$H_{0}$&    -3.412\\
& 6&   0.24600$-$0.34200   &211&    -3.710~$\pm$~0.615     &    -3.412&     0.484&     2.717&     0.178&Accept~$H_{0}$&    -3.465\\
& 7&   0.34200$-$0.65300   &270&    -2.428~$\pm$~0.191     &    -3.465&     5.421&     2.711&     2.000&Reject~$H_{0}$&    -1.390\\
$I$&      1&  -0.21900$-$-0.07000  &         207&    -3.891~$\pm$~0.407     &       ---&       ---&       ---&       ---&       ---         &       ---\\
& 2&  -0.07000$-$0.00100   &212&    -2.904~$\pm$~0.772     &    -3.891&     1.278&     2.716&     0.471&Accept~$H_{0}$&    -3.427\\
& 3&   0.00100$-$0.07100   &210&    -3.349~$\pm$~0.791     &    -3.427&     0.099&     2.717&     0.036&Accept~$H_{0}$&    -3.424\\
& 4&   0.07100$-$0.15000   &208&    -3.564~$\pm$~0.593     &    -3.424&     0.236&     2.717&     0.087&Accept~$H_{0}$&    -3.436\\
& 5&   0.15000$-$0.24600   &212&    -4.021~$\pm$~0.544     &    -3.436&     1.076&     2.716&     0.396&Accept~$H_{0}$&    -3.668\\
& 6&   0.24600$-$0.34200   &211&    -3.755~$\pm$~0.490     &    -3.668&     0.178&     2.717&     0.066&Accept~$H_{0}$&    -3.673\\
& 7&   0.34200$-$0.65300   &270&    -2.795~$\pm$~0.154     &    -3.673&     5.689&     2.711&     2.099&Reject~$H_{0}$&    -1.830\\
$J$&      1&   0.00200$-$0.16800   &         110&    -2.476~$\pm$~0.423     &       ---&       ---&       ---&       ---&       ---         &       ---\\
& 2&   0.16800$-$0.25600   &110&    -4.242~$\pm$~0.745     &    -2.476&     2.369&     2.621&     0.904&Accept~$H_{0}$&    -4.072\\
& 3&   0.25600$-$0.32600   &108&    -3.960~$\pm$~0.851     &    -4.072&     0.132&     2.622&     0.050&Accept~$H_{0}$&    -4.066\\
& 4&   0.32600$-$0.41800   &112&    -3.838~$\pm$~0.588     &    -4.066&     0.389&     2.620&     0.148&Accept~$H_{0}$&    -4.032\\
& 5&   0.41800$-$0.65300   &154&    -2.704~$\pm$~0.299     &    -4.032&     4.444&     2.608&     1.704&Reject~$H_{0}$&    -1.769\\
$H$&      1&   0.00200$-$0.16700   &         109&    -2.600~$\pm$~0.444     &       ---&       ---&       ---&       ---&       ---         &       ---\\
& 2&   0.16700$-$0.25500   &110&    -3.692~$\pm$~0.755     &    -2.600&     1.446&     2.621&     0.552&Accept~$H_{0}$&    -3.202\\
& 3&   0.25500$-$0.32500   &111&    -3.785~$\pm$~0.815     &    -3.202&     0.715&     2.621&     0.273&Accept~$H_{0}$&    -3.361\\
& 4&   0.32500$-$0.41500   &110&    -2.984~$\pm$~0.555     &    -3.361&     0.679&     2.621&     0.259&Accept~$H_{0}$&    -3.264\\
& 5&   0.41500$-$0.65300   &157&    -2.921~$\pm$~0.264     &    -3.264&     1.296&     2.608&     0.497&Accept~$H_{0}$&    -3.093\\
$K_{s}$&      1&   0.19000$-$0.26000   &          90&    -2.727~$\pm$~1.184     &       ---&       ---&       ---&       ---&       ---         &       ---\\
& 2&   0.26000$-$0.32000   & 90&    -5.291~$\pm$~1.097     &    -2.727&     2.337&     2.632&     0.888&Accept~$H_{0}$&    -5.004\\
& 3&   0.32000$-$0.38700   & 89&    -0.508~$\pm$~0.958     &    -5.004&     4.691&     2.632&     1.782&Reject~$H_{0}$&     3.008\\
$3.6\mu \mathrm{m}$&      1&  -0.22000$-$-0.06500  &         209&    -3.436~$\pm$~0.359     &       ---&       ---&       ---&       ---&       ---         &       ---\\
& 2&  -0.06500$-$0.00800   &211&    -2.501~$\pm$~0.677     &    -3.436&     1.381&     2.717&     0.509&Accept~$H_{0}$&    -2.960\\
& 3&   0.00800$-$0.07700   &207&    -3.714~$\pm$~0.665     &    -2.960&     1.132&     2.717&     0.417&Accept~$H_{0}$&    -3.274\\
& 4&   0.07700$-$0.16100   &212&    -4.383~$\pm$~0.481     &    -3.274&     2.307&     2.716&     0.849&Accept~$H_{0}$&    -4.216\\
& 5&   0.16100$-$0.25500   &209&    -4.074~$\pm$~0.443     &    -4.216&     0.320&     2.717&     0.118&Accept~$H_{0}$&    -4.199\\
& 6&   0.25500$-$0.36300   &212&    -3.588~$\pm$~0.322     &    -4.199&     1.898&     2.716&     0.699&Accept~$H_{0}$&    -3.772\\
& 7&   0.36300$-$0.65300   &238&    -3.204~$\pm$~0.148     &    -3.772&     3.835&     2.714&     1.413&Reject~$H_{0}$&    -2.970\\
$4.5\mu \mathrm{m}$&      1&  -0.22000$-$-0.06600  &         210&    -2.695~$\pm$~0.416     &       ---&       ---&       ---&       ---&       ---         &       ---\\
& 2&  -0.06600$-$0.00300   &209&    -2.120~$\pm$~0.827     &    -2.695&     0.695&     2.717&     0.256&Accept~$H_{0}$&    -2.548\\
& 3&   0.00300$-$0.07300   &211&    -3.710~$\pm$~0.812     &    -2.548&     1.432&     2.717&     0.527&Accept~$H_{0}$&    -3.160\\
& 4&   0.07300$-$0.15400   &210&    -4.139~$\pm$~0.575     &    -3.160&     1.701&     2.717&     0.626&Accept~$H_{0}$&    -3.773\\
& 5&   0.15400$-$0.24800   &209&    -3.898~$\pm$~0.512     &    -3.773&     0.245&     2.717&     0.090&Accept~$H_{0}$&    -3.784\\
& 6&   0.24800$-$0.34900   &211&    -4.420~$\pm$~0.423     &    -3.784&     1.503&     2.717&     0.553&Accept~$H_{0}$&    -4.136\\
& 7&   0.34900$-$0.65300   &268&    -3.211~$\pm$~0.140     &    -4.136&     6.593&     2.711&     2.432&Reject~$H_{0}$&    -1.886\\
$W_{V,I}$&      1&  -0.22000$-$-0.06700  &         210&    -3.955~$\pm$~0.289     &       ---&       ---&       ---&       ---&       ---         &       ---\\
& 2&  -0.06700$-$0.00200   &207&    -2.933~$\pm$~0.591     &    -3.955&     1.728&     2.717&     0.636&Accept~$H_{0}$&    -3.305\\
& 3&   0.00200$-$0.07300   &209&    -3.514~$\pm$~0.521     &    -3.305&     0.400&     2.717&     0.147&Accept~$H_{0}$&    -3.336\\
& 4&   0.07300$-$0.15500   &214&    -4.135~$\pm$~0.399     &    -3.336&     2.003&     2.716&     0.737&Accept~$H_{0}$&    -3.925\\
& 5&   0.15500$-$0.25000   &209&    -4.007~$\pm$~0.404     &    -3.925&     0.205&     2.717&     0.075&Accept~$H_{0}$&    -3.931\\
& 6&   0.25000$-$0.35100   &211&    -4.005~$\pm$~0.308     &    -3.931&     0.241&     2.717&     0.089&Accept~$H_{0}$&    -3.938\\
& 7&   0.35100$-$0.65300   &265&    -3.451~$\pm$~0.120     &    -3.938&     4.049&     2.711&     1.493&Reject~$H_{0}$&    -3.212\\
$V-I$&      1&  -0.21900$-$-0.06800  &         210&    -0.034~$\pm$~0.130     &       ---&       ---&       ---&       ---&       ---         &       ---\\
& 2&  -0.06800$-$0.00200   &210&    -0.046~$\pm$~0.209     &    -0.034&     0.056&     2.717&     0.021&Accept~$H_{0}$&    -0.034\\
& 3&   0.00200$-$0.07200   &207&     0.216~$\pm$~0.216     &    -0.034&     1.157&     2.717&     0.426&Accept~$H_{0}$&     0.072\\
& 4&   0.07200$-$0.15000   &210&     0.194~$\pm$~0.181     &     0.072&     0.668&     2.717&     0.246&Accept~$H_{0}$&     0.102\\
& 5&   0.15000$-$0.24700   &213&     0.151~$\pm$~0.148     &     0.102&     0.333&     2.716&     0.122&Accept~$H_{0}$&     0.108\\
& 6&   0.24700$-$0.34900   &210&     0.041~$\pm$~0.139     &     0.108&     0.480&     2.717&     0.177&Accept~$H_{0}$&     0.096\\
& 7&   0.34900$-$0.65300   &259&     0.394~$\pm$~0.047     &     0.096&     6.332&     2.712&     2.335&Reject~$H_{0}$&     0.791\\
\hline
\end{tabular}
\end{center}
{\footnotesize \textbf{Notes}: The  meaning of each column header - $n$ represents the number of non-overlapping subsets and 
$\log(P)$ is the period range in each subset. $N$ and $\hat{\beta}$ represent the number of stars and slope of linear 
regression in each subset, respectively. $\beta_{0}$ and $\beta_{w}$ represent the initial and updated testimator slope
after each hypothesis testing. $|t_{obs}|$ is estimated using errors on the slopes and the standard deviation of
PL relations and $t_{c}$ represents the theoretical T-value for confidence level of more than 95\%. $k$ is the 
probability of initial guess of the testimator being true and leads to the decision of acceptance/rejection. }
\end{minipage}
\end{table*}

Non-linearity in the PL relations for Cepheids in the LMC is discussed in detail in \citet{cpapir3}. The LMC FU mode 
Cepheid PL relations exhibit a break at 10 days at optical and near-infrared wavelengths and the FO mode Cepheids
provide evidence of a break in the PL relation at 2.5 days only at optical wavelengths. However, \citet{ngeow15} 
suggest there is no break in the SMC FU mode Cepheid PL relations at 10 days, if we do not consider stars with 
periods below 2.5 days. Further, \citet{subra15} found a break in the PL relations of both the FU and FO mode 
Cepheids at $P\sim2.95$ days and $P\sim1$ day, respectively. \citet{subra15} did not use any statistical tests to 
determine the significance of these observed breaks. We will test for any possible statistically significant 
non-linearity in SMC FO mode Cepheid PL relations using  robust statistical tests such as the F-test, random 
walk method and the testimator. These test statistics are discussed in detail in \citet{cpapir3} and will not 
be repeated here. In brief, the F-test compares a single regression line model over the entire period range with a two line 
regression, under the null hypothesis that the data follows a linear model. The random walk is a non-parametric test
and uses the partial sum of random permutation of residuals to test any departure from linearity. The testimator
compares the slopes of a number of subsets of the data to find evidence of change in the slope of the PL relation.

The results of the F-test, random walk and the testimator analysis are provided in Table~\ref{table:fo_smc_frw} and 
\ref{table:fo_smc_tm}. We test for non-linearity in the PL relations at 2.5 days and 1 day following the observed 
breaks in the PC \citep{bhardwaj14} and PL relations \citep{subra15}, respectively. At optical wavelengths, PL 
and PW relations provide evidence for a significant break at both 2.5 days and 1 day, according to all test statistics. 
The results of all the test statistics imply that we can reject the null hypothesis that a single regression line is 
a better model at a very high significance level. However, we do not find any significant change in the slope of 
PL relations at these periods in infrared bands. We use our tests to determine the significance of the break 
at 1.6 days, which is related to a feature in Fourier parameters for SMC FO mode Cepheids. The F-test and random 
walk find evidence of a significant break at 1.6 days in PL and PW relations. Independently, the testimator 
suggests a break in the PL and PW relations in a bin including 2.5 days period.

We note that \citet{subra15} found evidence of a break at 1 day. The assumed break point at 2.5 days and 1.6 days are related 
to non-linearities in PC relations at various phases of pulsation and also with sharp changes in light curve 
parameters for FO mode Cepheids \citep{bhardwaj14}. In order to find the most dominant break point in optical band
PL and PW relations, we apply period-cuts before using test statistics. If we consider stars with $P>1$~day in our
analysis, we still find significant change in the slope of the optical band relations at 2.5 days. However, if we only consider 
stars with $P<2.5$~days, we do not find evidence of a non-linearity at 1 day in $I$-band PL and the Wesenheit relation.
While the F-test suggests a break in the $V$-band PL relation, the random walk test does not support this non-linearity. 
Therefore, we emphasize that the non-linearity observed at 1 day in \citet{subra15} is influenced by the change in slope
of PL relations for P$>$ 2.5 days. Thus, the most dominant break in FO mode Cepheid PL relations at optical bands 
occurs at 2.5 days, similar to FU mode Cepheids.

\begin{figure}
\begin{center}
\includegraphics[width=0.5\textwidth,keepaspectratio]{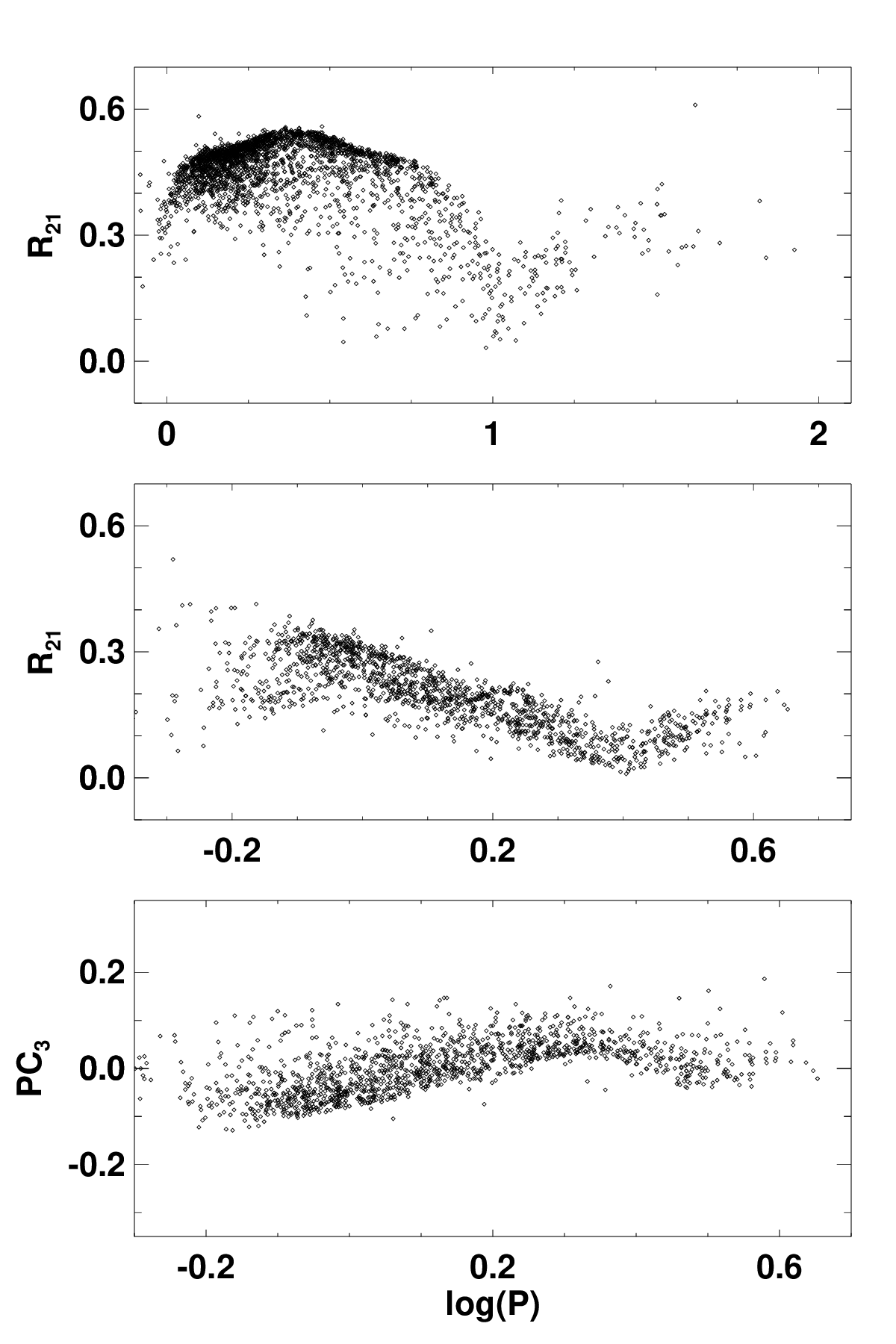}
\caption{Light curve parameters for FU and FO mode Cepheids in the SMC. The top panel displays the $I$-band Fourier amplitude parameter
\citep{bhardwaj15} for FU mode Cepheids. The middle panel shows Fourier amplitude parameter for FO mode Cepheids. The bottom
panel shows the variation of the third principal component \citep{deb09} of $I$-band light curves of SMC FO mode Cepheids.}
\label{fig:fou_pca}
\end{center}
\end{figure}

Fig.~\ref{fig:fou_pca} presents the light curve parameters for FU and FO mode Cepheids in the SMC. The top panel 
displays the $I$-band Fourier amplitude parameter \citep{bhardwaj15} for FU mode Cepheids. We observe a
maxima around 2.5 days in $R_{21}$ parameter as a function of period. The shorter period Cepheids ($P<2.5$~days)
in the SMC occupy a different location such that stars with lower periods have smaller amplitude ratio.
The middle panel displays the Fourier amplitude parameter for FO mode Cepheids. This parameter clearly
exhibits a sharp change in the progression with period around $\log P = 0.4$. There is also a break around
$P = 1.6$~days. All these changes in the progressions are also visible in other amplitude and phase 
parameters. The bottom panel presents the variation of the third principal component \citep{deb09} obtained from the $I$-band light curves 
of SMC FO mode Cepheids. Again, a change in the progression around $\log P = 0.4$ is distinctly observed
and similar but less pronounced features are seen in first and second principal components. Our test 
statistics results provide clear evidence that the observed non-linearities are related to changes
in light curve structure.

We also test the robustness of our test statistics results under various assumptions such as, homoscedasticity,
independent and identically distributed variables and normality of residuals. The observations of Cepheids
are independent of each other and the average of PL residuals do not show any significant trend as a 
function of period. Further, the mean of PL residuals is close to zero and the majority of
the residuals follow a normal distribution. We also use quantile-quantile (q-q) plots to find the outliers
at the extreme ends and tested whether our results are sensitive to these outliers.
The robustness of F-test under these assumptions is discussed in detail in \citet{ngeow15} and \citet{cpapir3}.
However, our results are robust to these assumptions and are supported by non-parametric random walk test.

\section{A Comparison with Short Period Fundamental-Mode Cepheids}

\begin{table}
\begin{minipage}{1.0\hsize}
\begin{center}
\caption{Multi-band PL relations for SMC FU mode Cepheids. \label{table:final_pl_fu}}
\begin{tabular}{|l|c|c|c|c|}
\hline
\hline
Band	&	PL Slope& 	PL ZP	&	$\sigma$& 	$N$\\
\hline
\hline
                 $V$&     -2.919$\pm$0.018     &     17.892$\pm$0.008     &      0.254&         2487\\
                 $I$&     -3.133$\pm$0.014     &     17.349$\pm$0.007     &      0.204&         2477\\
                 $J$&     -3.149$\pm$0.016     &     16.861$\pm$0.008     &      0.213&         2150\\
                 $H$&     -3.082$\pm$0.020     &     16.482$\pm$0.010     &      0.246&         2114\\
 $3.6\mu\mathrm{m}$&     -3.381$\pm$0.013     &     16.554$\pm$0.006     &      0.179&         2443\\
 $4.5\mu\mathrm{m}$&     -3.343$\pm$0.014     &     16.497$\pm$0.006     &      0.191&         2449\\
           $W_{V,I}$&     -3.468$\pm$0.010     &     16.501$\pm$0.005     &      0.142&         2466\\
\hline
\end{tabular}
\end{center}
\end{minipage}
\end{table}

\begin{table*}
\begin{minipage}{1.0\hsize}
\begin{center}
\caption{Results of F and random walk tests for PL and PW and PC relations for FU mode Cepheids in the SMC to test non-linearity 
at various periods. The meaning of each column header is discussed in Table~\ref{table:fo_smc_frw}.}
\label{table:fu_smc_frw}
\begin{tabular}{|l|c|c|c|c|c|c|c|c|c|c|c|}
\hline
\hline
$Y$& PL Slope$_{S}$& PL ZP$_{S}$& $\sigma_{S}$& $N_{S}$ & PL Slope$_{L}$& PL ZP$_{L}$& $\sigma_{S}$& $N_{S}$ & $F$ & $p(F)$& $p(R)$\\
\hline
\hline
\multicolumn{12}{c}{For all $P$ with a break at 2.5 days \citep{bhardwaj14}}\\
\hline
                 $V$&    -3.057$\pm$0.064     &    17.935$\pm$0.015     &     0.243&        1567&    -2.678$\pm$0.035     &    17.704$\pm$0.025     &     0.261&         920&    34.865&     0.000&     0.000\\
                 $I$&    -3.267$\pm$0.052     &    17.389$\pm$0.012     &     0.195&        1557&    -2.930$\pm$0.028     &    17.192$\pm$0.020     &     0.208&         920&    38.858&     0.000&     0.000\\
                 $J$&    -2.732$\pm$0.068     &    16.783$\pm$0.016     &     0.216&        1246&    -3.059$\pm$0.028     &    16.779$\pm$0.020     &     0.199&         904&    32.633&     0.000&     0.000\\
                 $H$&    -2.281$\pm$0.086     &    16.307$\pm$0.021     &     0.266&        1197&    -3.156$\pm$0.030     &    16.523$\pm$0.021     &     0.202&         917&    54.788&     0.000&     0.000\\
         $3.6\mu \mathrm{m}$&    -3.519$\pm$0.049     &    16.588$\pm$0.011     &     0.185&        1532&    -3.242$\pm$0.024     &    16.449$\pm$0.017     &     0.164&         911&    21.795&     0.000&     0.000\\
         $4.5\mu \mathrm{m}$&    -3.375$\pm$0.052     &    16.513$\pm$0.012     &     0.199&        1531&    -3.196$\pm$0.025     &    16.382$\pm$0.018     &     0.172&         918&    20.144&     0.000&     0.000\\
           $W_{V,I}$&    -3.588$\pm$0.037     &    16.534$\pm$0.008     &     0.136&        1542&    -3.323$\pm$0.020     &    16.388$\pm$0.014     &     0.144&         924&    41.961&     0.000&     0.000\\
               $V-I$&     0.206$\pm$0.016     &     0.549$\pm$0.004     &     0.060&        1562&     0.272$\pm$0.009     &     0.503$\pm$0.006     &     0.066&         911&    23.729&     0.000&     0.000\\
\hline
\multicolumn{12}{c}{For all $P$ with a break at 2.95 days \citep{subra15}}\\
\hline
                 $V$&    -3.176$\pm$0.052     &    17.955$\pm$0.013     &     0.245&        1731&    -2.689$\pm$0.040     &    17.714$\pm$0.031     &     0.262&         756&    30.706&     0.000&     0.000\\
                 $I$&    -3.352$\pm$0.042     &    17.403$\pm$0.011     &     0.196&        1719&    -2.932$\pm$0.032     &    17.194$\pm$0.024     &     0.208&         758&    35.979&     0.000&     0.000\\
                 $J$&    -3.015$\pm$0.053     &    16.836$\pm$0.014     &     0.219&        1406&    -3.063$\pm$0.032     &    16.784$\pm$0.024     &     0.197&         744&     9.500&     0.000&     0.000\\
                 $H$&    -2.621$\pm$0.067     &    16.372$\pm$0.018     &     0.266&        1360&    -3.165$\pm$0.033     &    16.530$\pm$0.025     &     0.197&         754&    31.238&     0.000&     0.000\\
         $3.6\mu \mathrm{m}$&    -3.560$\pm$0.039     &    16.595$\pm$0.010     &     0.183&        1694&    -3.250$\pm$0.027     &    16.457$\pm$0.020     &     0.163&         749&    21.470&     0.000&     0.000\\
         $4.5\mu \mathrm{m}$&    -3.455$\pm$0.041     &    16.526$\pm$0.011     &     0.197&        1693&    -3.195$\pm$0.028     &    16.382$\pm$0.021     &     0.170&         756&    16.467&     0.000&     0.000\\
           $W_{V,I}$&    -3.641$\pm$0.030     &    16.543$\pm$0.008     &     0.137&        1705&    -3.324$\pm$0.022     &    16.390$\pm$0.017     &     0.144&         761&    39.886&     0.000&     0.000\\
               $V-I$&     0.175$\pm$0.013     &     0.554$\pm$0.003     &     0.061&        1724&     0.269$\pm$0.010     &     0.505$\pm$0.008     &     0.066&         749&    19.303&     0.000&     0.000\\
\hline
\end{tabular}
\end{center}
\end{minipage}
\end{table*}

\begin{table*}
\begin{minipage}{1.0\hsize}
\begin{center}
\caption{Results of the testimator on PL, PW and PC relations for FU mode Cepheids in the SMC.
The meaning of each column header is discussed in Table~\ref{table:fo_smc_tm}.}
\label{table:fu_smc_tm}
\begin{tabular}{|l|c|c|c|c|c|c|c|c|c|c|}
\hline
\hline
Band &$n$  &  $\log(P)$&	$N$&	$\hat{\beta}$&	$\beta_{0}$&	$|t_{obs}|$&	$t_{c}$&	$k$&	Decision&	$\beta_{w}$\\
\hline
\hline
$V$&      1&   0.00500$-$0.11100   &         269&    -3.178~$\pm$~0.530     &       ---&       ---&       ---&       ---&       ---         &       ---\\
& 2&   0.11100$-$0.16200   &269&    -3.890~$\pm$~1.014     &    -3.178&     0.703&     2.795&     0.251&Accept~$H_{0}$&    -3.357\\
& 3&   0.16200$-$0.20700   &267&    -3.910~$\pm$~1.030     &    -3.357&     0.537&     2.796&     0.192&Accept~$H_{0}$&    -3.463\\
& 4&   0.20700$-$0.25700   &274&    -2.627~$\pm$~1.001     &    -3.463&     0.835&     2.795&     0.299&Accept~$H_{0}$&    -3.213\\
& 5&   0.25700$-$0.31800   &270&    -5.124~$\pm$~0.780     &    -3.213&     2.449&     2.795&     0.876&Accept~$H_{0}$&    -4.887\\
& 6&   0.31800$-$0.42100   &271&    -4.628~$\pm$~0.565     &    -4.887&     0.459&     2.795&     0.164&Accept~$H_{0}$&    -4.845\\
& 7&   0.42100$-$0.53300   &270&    -2.520~$\pm$~0.491     &    -4.845&     4.732&     2.795&     1.693&Reject~$H_{0}$&    -0.909\\
$I$&      1&   0.00500$-$0.11100   &         268&    -3.539~$\pm$~0.449     &       ---&       ---&       ---&       ---&       ---         &       ---\\
& 2&   0.11100$-$0.16300   &270&    -4.097~$\pm$~0.813     &    -3.539&     0.686&     2.795&     0.245&Accept~$H_{0}$&    -3.676\\
& 3&   0.16300$-$0.20800   &271&    -3.581~$\pm$~0.819     &    -3.676&     0.116&     2.795&     0.042&Accept~$H_{0}$&    -3.672\\
& 4&   0.20800$-$0.26000   &271&    -2.089~$\pm$~0.792     &    -3.672&     1.998&     2.795&     0.715&Accept~$H_{0}$&    -2.540\\
& 5&   0.26000$-$0.32200   &270&    -4.336~$\pm$~0.614     &    -2.540&     2.923&     2.795&     1.046&Reject~$H_{0}$&    -4.418\\
$J$&      1&  -0.02000$-$0.14000   &         227&    -1.474~$\pm$~0.352     &       ---&       ---&       ---&       ---&       ---         &       ---\\
& 2&   0.14000$-$0.19000   &226&    -2.402~$\pm$~0.951     &    -1.474&     0.976&     2.800&     0.349&Accept~$H_{0}$&    -1.797\\
& 3&   0.19000$-$0.23600   &235&    -2.807~$\pm$~1.024     &    -1.797&     0.986&     2.799&     0.352&Accept~$H_{0}$&    -2.153\\
& 4&   0.23600$-$0.28700   &229&    -3.820~$\pm$~0.914     &    -2.153&     1.823&     2.799&     0.651&Accept~$H_{0}$&    -3.238\\
& 5&   0.28700$-$0.36300   &233&    -3.118~$\pm$~0.690     &    -3.238&     0.175&     2.799&     0.062&Accept~$H_{0}$&    -3.231\\
& 6&   0.36300$-$0.46000   &228&    -4.394~$\pm$~0.480     &    -3.231&     2.423&     2.800&     0.866&Accept~$H_{0}$&    -4.238\\
& 7&   0.46000$-$0.56100   &232&    -2.070~$\pm$~0.445     &    -4.238&     4.867&     2.799&     1.739&Reject~$H_{0}$&    -0.468\\
$H$&      1&  -0.02000$-$0.14700   &         230&    -1.235~$\pm$~0.528     &       ---&       ---&       ---&       ---&       ---         &       ---\\
& 2&   0.14700$-$0.19800   &229&     0.327~$\pm$~1.329     &    -1.235&     1.176&     2.799&     0.420&Accept~$H_{0}$&    -0.579\\
& 3&   0.19800$-$0.24400   &228&    -2.564~$\pm$~1.330     &    -0.579&     1.492&     2.800&     0.533&Accept~$H_{0}$&    -1.637\\
& 4&   0.24400$-$0.30000   &231&    -3.000~$\pm$~0.944     &    -1.637&     1.444&     2.799&     0.516&Accept~$H_{0}$&    -2.340\\
& 5&   0.30000$-$0.38300   &231&    -2.331~$\pm$~0.615     &    -2.340&     0.015&     2.799&     0.005&Accept~$H_{0}$&    -2.340\\
& 6&   0.38300$-$0.47800   &231&    -4.168~$\pm$~0.544     &    -2.340&     3.360&     2.799&     1.200&Reject~$H_{0}$&    -4.535\\
$3.6\mu \mathrm{m}$&      1&  -0.07500$-$0.11000   &         270&    -3.374~$\pm$~0.355     &       ---&       ---&       ---&       ---&       ---         &       ---\\
& 2&   0.11000$-$0.16300   &270&    -4.051~$\pm$~0.744     &    -3.374&     0.909&     2.795&     0.325&Accept~$H_{0}$&    -3.594\\
& 3&   0.16300$-$0.21000   &265&    -4.005~$\pm$~0.769     &    -3.594&     0.535&     2.796&     0.191&Accept~$H_{0}$&    -3.673\\
& 4&   0.21000$-$0.26400   &272&    -2.995~$\pm$~0.729     &    -3.673&     0.929&     2.795&     0.333&Accept~$H_{0}$&    -3.448\\
& 5&   0.26400$-$0.33000   &272&    -3.418~$\pm$~0.556     &    -3.448&     0.054&     2.795&     0.019&Accept~$H_{0}$&    -3.447\\
& 6&   0.33000$-$0.44200   &270&    -4.106~$\pm$~0.351     &    -3.447&     1.876&     2.795&     0.671&Accept~$H_{0}$&    -3.889\\
& 7&   0.44200$-$0.55400   &270&    -2.640~$\pm$~0.320     &    -3.889&     3.909&     2.795&     1.398&Reject~$H_{0}$&    -2.143\\
$4.5\mu \mathrm{m}$&      1&  -0.08100$-$0.11000   &         267&    -2.933~$\pm$~0.355     &       ---&       ---&       ---&       ---&       ---         &       ---\\
& 2&   0.11000$-$0.16300   &269&    -4.054~$\pm$~0.796     &    -2.933&     1.408&     2.795&     0.504&Accept~$H_{0}$&    -3.497\\
& 3&   0.16300$-$0.21000   &269&    -3.554~$\pm$~0.865     &    -3.497&     0.066&     2.795&     0.024&Accept~$H_{0}$&    -3.499\\
& 4&   0.21000$-$0.26400   &272&    -2.974~$\pm$~0.774     &    -3.499&     0.678&     2.795&     0.243&Accept~$H_{0}$&    -3.371\\
& 5&   0.26400$-$0.33000   &271&    -3.950~$\pm$~0.607     &    -3.371&     0.954&     2.795&     0.341&Accept~$H_{0}$&    -3.569\\
& 6&   0.33000$-$0.44200   &270&    -4.028~$\pm$~0.369     &    -3.569&     1.247&     2.795&     0.446&Accept~$H_{0}$&    -3.774\\
& 7&   0.44200$-$0.55400   &272&    -2.797~$\pm$~0.340     &    -3.774&     2.878&     2.795&     1.029&Reject~$H_{0}$&    -2.768\\
$W_{V,I}$&      1&   0.00500$-$0.11300   &         270&    -3.933~$\pm$~0.328     &       ---&       ---&       ---&       ---&       ---         &       ---\\
& 2&   0.11300$-$0.16500   &266&    -3.826~$\pm$~0.586     &    -3.933&     0.182&     2.796&     0.065&Accept~$H_{0}$&    -3.926\\
& 3&   0.16500$-$0.21100   &273&    -3.792~$\pm$~0.600     &    -3.926&     0.223&     2.795&     0.080&Accept~$H_{0}$&    -3.915\\
& 4&   0.21100$-$0.26300   &269&    -3.557~$\pm$~0.503     &    -3.915&     0.711&     2.795&     0.254&Accept~$H_{0}$&    -3.824\\
& 5&   0.26300$-$0.32600   &272&    -4.072~$\pm$~0.423     &    -3.824&     0.586&     2.795&     0.210&Accept~$H_{0}$&    -3.876\\
& 6&   0.32600$-$0.43800   &270&    -3.939~$\pm$~0.273     &    -3.876&     0.231&     2.795&     0.082&Accept~$H_{0}$&    -3.881\\
& 7&   0.43800$-$0.54500   &269&    -2.892~$\pm$~0.302     &    -3.881&     3.274&     2.795&     1.171&Reject~$H_{0}$&    -2.723\\
$V-I$&      1&   0.00500$-$0.11100   &         269&     0.116~$\pm$~0.119     &       ---&       ---&       ---&       ---&       ---         &       ---\\
& 2&   0.11100$-$0.16200   &271&     0.379~$\pm$~0.234     &     0.116&     1.126&     2.795&     0.403&Accept~$H_{0}$&     0.222\\
& 3&   0.16200$-$0.20700   &267&    -0.051~$\pm$~0.261     &     0.222&     1.044&     2.796&     0.374&Accept~$H_{0}$&     0.120\\
& 4&   0.20700$-$0.25800   &273&     0.440~$\pm$~0.271     &     0.120&     1.180&     2.795&     0.422&Accept~$H_{0}$&     0.255\\
& 5&   0.25800$-$0.32100   &269&    -0.078~$\pm$~0.210     &     0.255&     1.588&     2.795&     0.568&Accept~$H_{0}$&     0.066\\
& 6&   0.32100$-$0.42500   &271&    -0.276~$\pm$~0.131     &     0.066&     2.600&     2.795&     0.930&Accept~$H_{0}$&    -0.252\\
& 7&   0.42500$-$0.53800   &270&     0.375~$\pm$~0.116     &    -0.252&     5.398&     2.795&     1.931&Reject~$H_{0}$&     0.958\\
\hline
\end{tabular}
\end{center}
\end{minipage}
\end{table*}

\citet{ngeow15} derived the multi-band PL relations for SMC FU mode Cepheids with $P>2.5$~days. 
Therefore, in our analysis we include the short period Cepheids to compare with FO mode
Cepheids and simultaneously test for any evidence of a possible non-linearity in PL, PW and PC 
relations at short periods. The short period breaks of around 2 days in the PL relation for FU mode SMC Cepheids 
was first reported by \citet{bauer99} and further revisited by \citet[][at 2.5 days]{tammann09}. 
\citet{bhardwaj14} discussed these breaks in PC \& A-C relations as a function of pulsation phase at 2.5 days.
However, \citet{subra15} found a break in PL and PC relations for FU mode Cepheids at 2.95 days.
In our analysis, we make use of OGLE-III counterparts for FU mode Cepheids in 2MASS and SAGE-SMC catalogs, taken
from \citet{ngeow15}. We use extinction corrected mean magnitudes and apply 2.5$\sigma$ clipping to fit PL, 
PW and PC relations. The results of a single regression line over the entire period range are provided in
Table~\ref{table:final_pl_fu}. We note that these PL relations are significantly different to \citet{ngeow15} 
results, presumably due to inclusion of short period ($P<2.5$~days) Cepheids.

\begin{figure}
\begin{center}
\includegraphics[width=0.4\textwidth,keepaspectratio]{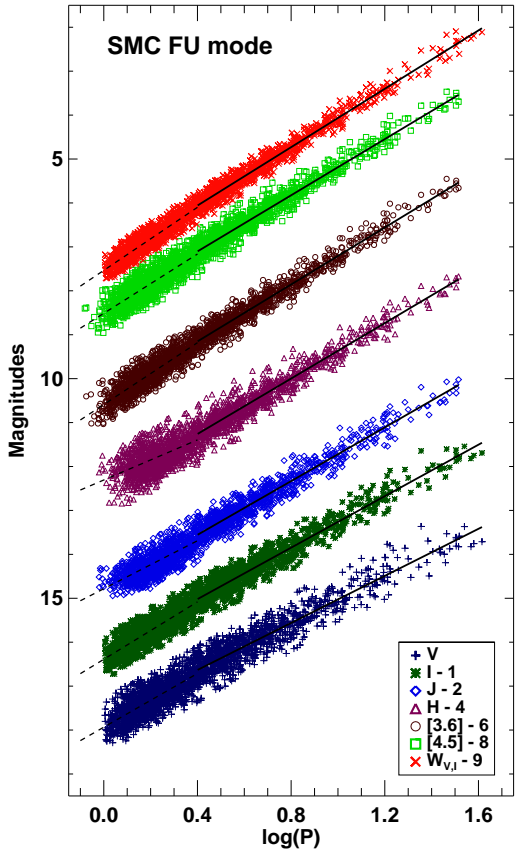}
\caption{Multi-band PL relations and optical Wesenheit relation for SMC FU mode Cepheids.
The solid/dashed lines represent the best fit regression for Cepheids with periods below and above 2.5 days.}
\label{fig:final_pl_fu}
\end{center}
\end{figure}

\begin{figure}
\begin{center}
\includegraphics[width=0.4\textwidth,keepaspectratio]{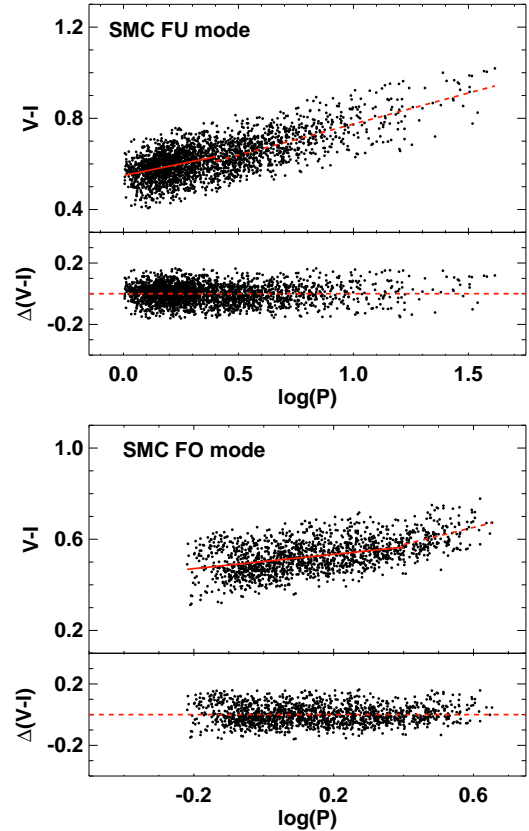}
\caption{Optical-band PC relation for SMC FU and FO mode Cepheids. The solid/dashed lines 
represent the best fit regression for Cepheids with periods below and above 2.5 days. 
The residuals from a single regression line are shown in the bottom panels. }
\label{fig:vi_col}
\end{center}
\end{figure}

The plots of multi-band PL relations and optical Wesenheit for FU mode Cepheids are displayed in Fig.~\ref{fig:final_pl_fu}. 
The results of the F-test, Random Walk and the testimator analysis are provided in Table~\ref{table:fu_smc_frw} 
and \ref{table:fu_smc_tm}, respectively. \citet{ngeow15} found that the $K_s$, $5.8$ \& $8.0\mu\mathrm{m}$ bands PL
relations are influenced by the incompleteness at short period ends and, therefore, required a period cut. 
Hence, we do not use these bands in our study of short period breaks in PL relations.
The optical $V$- and $I$-band PL relations and the Wesenheit function for SMC FU mode Cepheids provide 
evidence for a significant variation in slope between Cepheids having periods smaller/greater than 2.5 days. 
Similarly, the near-infrared ($J$ \& $H$) and mid-infrared ($3.6~\&~4.5\mu\mathrm{m}$) band PL relations also 
provide evidence for a very large deviation in slope for short and long period range Cepheids. 
Visual inspection of $H$-band PL relation from \citet{ngeow15} suggests that the magnitudes may be
approaching incompleteness limit at the short period end leading to a large deviation in slope. However, the
magnitudes in other infrared bands lie well above their detection limit and, therefore, any evidence of 
non-linearity is not influenced by the incompleteness bias. We also use our test statistics to determine the 
significance of the break at 2.95 days as observed in \citet{subra15}. Both F-test and Random Walk imply a significant 
change in the slope of the multi-band PL relations at this period as well. Interestingly, the testimator also finds a 
change in the slope of optical band PL and Wesenheit relations in the period bin containing 2.95 days.

The plots of optical band PC relations for FU and FO mode Cepheids are shown in Fig.~\ref{fig:vi_col}. 
The results of the test statistics are provided in Table~\ref{table:fo_smc_frw} \& \ref{table:fu_smc_frw}
for F and random walk test and in Table~\ref{table:fo_smc_tm} \& ~\ref{table:fu_smc_tm} for the testimator.
All test statistics provide evidence of a significant change in the slope of PC relations 
for both FU and FO mode Cepheids having periods smaller/greater than 2.5 days. We have shown that the
SMC FO mode PL relations do not provide evidence of a break at 1 day as suggested by \citet{subra15}, 
if we restrict the sample size to $P<2.5$~days. However, we do find a significant change in the slope of 
PC relation at 1 day for this case as well. There is also a significant deviation in the slope of PC at 2.95 days 
for FU mode Cepheids.

\section{COMPARISON WITH LMC PL RELATIONS}

We also compare our PL relations for SMC FO mode Cepheids with PL relations for LMC FO mode Cepheids. At optical 
wavelengths, the OGLE-III catalog has classified 1238 FO mode Cepheids in the LMC. We note that \citet{cpapir3} provided
variation in the slope of PL relations for FO mode Cepheids in the LMC and tested for possible non-linearity. Therefore, we
derive PL and PW relations for these FO mode Cepheids with similar period cut and sigma clipping as carried out for the FO 
mode Cepheids in the SMC, for relative comparison. Recently, \citet{macri15} derived new near-infrared 
PL relations for the LMC FO mode Cepheids 
based on time series observations. Therefore, we do not use 2MASS data to derive $JHK_s$ PL relations as \citet{macri15} 
results are clearly superior in all aspects. These PL relations at $VIJHK_s$ wavelengths are tested to determine
possible non-linearities in multiple wavelengths \citep{cpapir3}. We note that a detailed study on 
mid-infrared PL relations does not exist in literature for FO mode Cepheids. Therefore, we derive new LMC FO 
Cepheid PL relations at mid-infrared wavelengths for OGLE-III Cepheids.

\begin{table}
\begin{minipage}{1.0\hsize}
\begin{center}
\caption{Summary of the matched OGLE-III LMC FO mode Cepheids with SAGE archival data. 
The meaning of each column header is discussed in Table~\ref{table:irac_delta}. \label{table:irac_delta1}}
\begin{tabular}{|l|c|c|c|c|c|}
\hline
\hline
Band  &  epoch& $N_{\mathrm{match}}$ & $<\Delta>^{a}$ & $\sigma^b$ & $^c$(in \%) \\
\hline
\hline
3.6$\mu\mathrm{m}$& 1&          371&      0.223&     0.225&      97.84\\
          & 2&          818&      0.129&     0.171&      99.02\\
4.5$\mu\mathrm{m}$& 1&          364&      0.220&     0.214&      98.08\\
          & 2&          810&      0.128&     0.171&      99.14\\
5.8$\mu\mathrm{m}$& 1&          204&      0.208&     0.208&      98.04\\
          & 2&          464&      0.126&     0.178&      99.14\\
8.0$\mu\mathrm{m}$& 1&           69&      0.278&     0.329&      94.20\\
          & 2&          176&      0.137&     0.237&      97.73\\
\hline
\end{tabular}
\end{center}
\end{minipage}
\end{table}

\begin{table}
\begin{minipage}{1.0\hsize}
\begin{center}
\caption{Optical and mid-infrared band PL relations for LMC FO mode Cepheids. \label{table:lmc_mir}}
\begin{tabular}{|l|c|c|c|c|}
\hline
\hline
Band	&	PL Slope& 	PL ZP	&	$\sigma$& 	$N$\\
\hline
\hline
                   V&     -3.299~$\pm$~0.029     &     16.865~$\pm$~0.010     &      0.182&         1084\\
                   I&     -3.354~$\pm$~0.021     &     16.294~$\pm$~0.007     &      0.129&         1084\\
           $W_{V,I}$&     -3.466~$\pm$~0.011     &     15.415~$\pm$~0.004     &      0.069&         1075\\
         $3.6\mu\mathrm{m}$&     -3.445~$\pm$~0.017     &     15.490~$\pm$~0.006     &      0.099&         1055\\
         $4.5\mu\mathrm{m}$&     -3.424~$\pm$~0.018     &     15.460~$\pm$~0.007     &      0.104&         1049\\
\hline
\end{tabular}
\end{center}
\end{minipage}
\end{table}

\begin{figure}
\begin{center}
\includegraphics[width=0.49\textwidth,keepaspectratio]{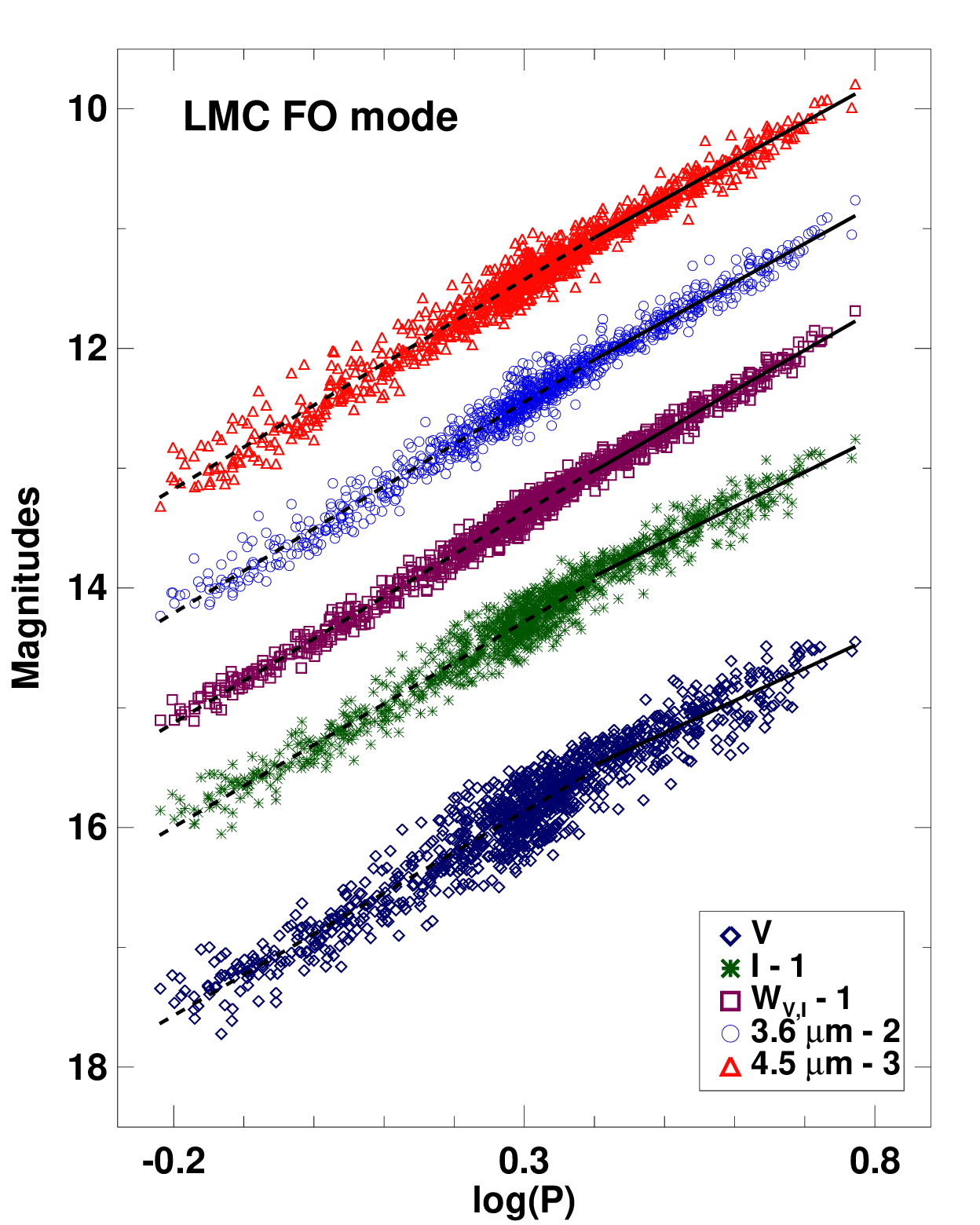}
\caption{Multi band PL relations and Wesenheit function for LMC FO mode Cepheids. 
The solid/dashed lines represent the best fit regression
for Cepheids with periods below and above 2.5 days.}
\label{fig:lmc_mir}
\end{center}
\end{figure}

\begin{table*}
\begin{minipage}{1.0\hsize}
\begin{center}
\caption{Results of F and random walk tests for PL and PW relations for FO mode Cepheids in the LMC to test non-linearity 
at 2.5 days. The meaning of each column header is discussed in Table~\ref{table:fo_smc_frw}.}
\label{table:fo_lmc_frw}
\begin{tabular}{|l|c|c|c|c|c|c|c|c|c|c|c|}
\hline
\hline
$Y$& PL Slope$_{S}$& PL ZP$_{S}$& $\sigma_{S}$& $N_{S}$ & PL Slope$_{L}$& PL ZP$_{L}$& $\sigma_{S}$& $N_{S}$ & $F$ & $p(F)$& $p(R)$\\
\hline
\hline
\multicolumn{12}{c}{For all $P$ with a break at 2.5 days \citep{bhardwaj14}}\\
\hline
                 $V$&    -3.410$\pm$0.045     &    16.888$\pm$0.012     &     0.187&         802&    -2.688$\pm$0.108     &    16.554$\pm$0.057     &     0.155&         282&    16.080&     0.000&     0.000\\
                 $I$&    -3.430$\pm$0.032     &    16.310$\pm$0.008     &     0.132&         802&    -2.894$\pm$0.078     &    16.058$\pm$0.041     &     0.112&         282&    17.231&     0.000&     0.000\\
  $3.6\mu\mathrm{m}$&    -3.524$\pm$0.027     &    15.504$\pm$0.007     &     0.104&         769&    -3.231$\pm$0.057     &    15.386$\pm$0.030     &     0.079&         286&    11.088&     0.000&     0.000\\
  $4.5\mu\mathrm{m}$&    -3.499$\pm$0.030     &    15.474$\pm$0.008     &     0.112&         764&    -3.223$\pm$0.055     &    15.364$\pm$0.029     &     0.077&         285&     8.845&     0.000&     0.001\\
           $W_{V,I}$&    -3.517$\pm$0.017     &    15.424$\pm$0.005     &     0.070&         790&    -3.359$\pm$0.045     &    15.366$\pm$0.024     &     0.066&         285&     9.145&     0.000&     0.000\\
\hline
\end{tabular}
\end{center}
\end{minipage}
\end{table*}

\begin{table*}
\begin{minipage}{1.0\hsize}
\begin{center}
\caption{Results of the testimator on PL and PW relations for FO mode Cepheids in the LMC. The meaning of each 
column header is discussed in Table~\ref{table:fo_smc_tm}.}
\label{table:fo_lmc_tm}
\begin{tabular}{|l|c|c|c|c|c|c|c|c|c|c|}
\hline
\hline
Band &$n$  &  $\log(P)$&	$N$&	$\hat{\beta}$&	$\beta_{0}$&	$|t_{obs}|$&	$t_{c}$&	$k$&	Decision&	$\beta_{w}$\\
\hline
\hline
$V$&      1&  -0.21900$-$0.12700   &         180&    -2.997$\pm$0.133     &       ---&       ---&       ---&       ---&       ---         &       ---\\
& 2&   0.12700$-$0.26400   &172&    -4.392$\pm$0.419     &    -2.997&     3.329&     2.669&     1.247&Reject~$H_{0}$&    -4.737\\
$I$&      1&  -0.21900$-$0.12700   &         180&    -3.045$\pm$0.101     &       ---&       ---&       ---&       ---&       ---         &       ---\\
& 2&   0.12700$-$0.26500   &179&    -3.707$\pm$0.284     &    -3.045&     2.328&     2.668&     0.872&Accept~$H_{0}$&    -3.623\\
& 3&   0.26500$-$0.31000   &179&    -3.138$\pm$0.790     &    -3.623&     0.613&     2.668&     0.230&Accept~$H_{0}$&    -3.511\\
& 4&   0.31000$-$0.36400   &181&    -3.467$\pm$0.607     &    -3.511&     0.074&     2.668&     0.028&Accept~$H_{0}$&    -3.510\\
& 5&   0.36400$-$0.47000   &181&    -3.386$\pm$0.257     &    -3.510&     0.482&     2.668&     0.181&Accept~$H_{0}$&    -3.488\\
& 6&   0.47000$-$0.77200   &183&    -2.887$\pm$0.133     &    -3.488&     4.532&     2.667&     1.699&Reject~$H_{0}$&    -2.466\\
$3.6\mu\mathrm{m}$&      1&  -0.21900$-$0.14100   &         170&    -3.251$\pm$0.099     &       ---&       ---&       ---&       ---&       ---         &       ---\\
& 2&   0.14100$-$0.26700   &168&    -3.434$\pm$0.229     &    -3.251&     0.799&     2.670&     0.299&Accept~$H_{0}$&    -3.306\\
& 3&   0.26700$-$0.31000   &172&    -4.158$\pm$0.717     &    -3.306&     1.189&     2.669&     0.445&Accept~$H_{0}$&    -3.685\\
& 4&   0.31000$-$0.36000   &169&    -3.484$\pm$0.527     &    -3.685&     0.381&     2.670&     0.143&Accept~$H_{0}$&    -3.657\\
& 5&   0.36000$-$0.45300   &171&    -3.376$\pm$0.217     &    -3.657&     1.291&     2.669&     0.483&Accept~$H_{0}$&    -3.521\\
& 6&   0.45300$-$0.77200   &204&    -3.195$\pm$0.082     &    -3.521&     3.962&     2.664&     1.487&Reject~$H_{0}$&    -3.036\\
$4.5\mu\mathrm{m}$&      1&  -0.21900$-$0.14300   &         170&    -3.317$\pm$0.119     &       ---&       ---&       ---&       ---&       ---         &       ---\\
& 2&   0.14300$-$0.26800   &169&    -3.560$\pm$0.252     &    -3.317&     0.961&     2.670&     0.360&Accept~$H_{0}$&    -3.405\\
& 3&   0.26800$-$0.31000   &169&    -4.685$\pm$0.737     &    -3.405&     1.738&     2.670&     0.651&Accept~$H_{0}$&    -4.238\\
& 4&   0.31000$-$0.36200   &172&    -3.484$\pm$0.480     &    -4.238&     1.572&     2.669&     0.589&Accept~$H_{0}$&    -3.794\\
& 5&   0.36200$-$0.45700   &167&    -3.595$\pm$0.240     &    -3.794&     0.829&     2.670&     0.310&Accept~$H_{0}$&    -3.732\\
& 6&   0.45700$-$0.77200   &201&    -3.177$\pm$0.080     &    -3.732&     6.908&     2.665&     2.592&Reject~$H_{0}$&    -2.294\\
$W_{V,I}$&      1&  -0.21900$-$0.12800   &         170&    -3.302$\pm$0.059     &       ---&       ---&       ---&       ---&       ---         &       ---\\
& 2&   0.12800$-$0.26400   &165&    -3.535$\pm$0.161     &    -3.302&     1.452&     2.670&     0.544&Accept~$H_{0}$&    -3.428\\
& 3&   0.26400$-$0.30600   &174&    -3.795$\pm$0.483     &    -3.428&     0.759&     2.669&     0.284&Accept~$H_{0}$&    -3.533\\
& 4&   0.30600$-$0.35300   &169&    -3.483$\pm$0.352     &    -3.533&     0.140&     2.670&     0.052&Accept~$H_{0}$&    -3.530\\
& 5&   0.35300$-$0.44000   &170&    -3.357$\pm$0.172     &    -3.530&     1.007&     2.669&     0.377&Accept~$H_{0}$&    -3.465\\
& 6&   0.44000$-$0.77200   &226&    -3.383$\pm$0.058     &    -3.465&     1.406&     2.662&     0.528&Accept~$H_{0}$&    -3.422\\
\hline
\end{tabular}
\end{center}
\end{minipage}
\end{table*}

\begin{table*}
\begin{minipage}{1.0\hsize}
\begin{center}
\caption{Comparison of the slopes of FO mode Cepheid PL relations between the SMC and LMC. \label{table:comp_slopes}}
\begin{tabular}{|l|c|c|c|c|c|c|c|c|}
\hline
\hline
Band	&Galaxy&	PL Slope& 	$\sigma$& 	$N$&	$|\Delta|^a~\pm~\sigma_{\Delta}^b$&	Src&	$|T|$&	$p(t)$\\
\hline
\hline
       $V$&SMC&     -3.147$\pm$0.035     &     0.268&   1531&     0.152$\pm$0.045     &TW&     3.124&     0.002\\
          &LMC&     -3.299$\pm$0.029     &     0.182&   1084&                        &B15&       ---&       ---\\
       $I$&SMC&     -3.332$\pm$0.028     &     0.218&   1531&     0.022$\pm$0.035     &TW&     0.569&     0.569\\
          &LMC&     -3.354$\pm$0.021     &     0.129&   1084&                        &B15&       ---&       ---\\
       $J$&SMC&     -3.095$\pm$0.055     &     0.198&    595&     0.202$\pm$0.059     &TW&     3.859&     0.000\\
          &LMC&     -3.297$\pm$0.020     &     0.134&    475&                        &M15&       ---&       ---\\
       $H$&SMC&     -3.151$\pm$0.053     &     0.191&    598&     0.064$\pm$0.057     &TW&     1.319&     0.187\\
          &LMC&     -3.215$\pm$0.020     &     0.100&    475&                        &M15&       ---&       ---\\
     $K_s$&SMC&     -3.132$\pm$0.083     &     0.197&    457&     0.113$\pm$0.086     &TW&     1.579&     0.115\\
          &LMC&     -3.245$\pm$0.023     &     0.086&    475&                        &M15&       ---&       ---\\
$3.6\mu\mathrm{m}$&SMC&     -3.510$\pm$0.024     &     0.180&   1499&     0.065$\pm$0.029     &TW&     1.969&     0.049\\
          &LMC&     -3.445$\pm$0.017     &     0.099&   1055&                        &TW&       ---&       ---\\
$4.5\mu\mathrm{m}$&SMC&     -3.451$\pm$0.028     &     0.200&   1529&     0.027$\pm$0.033     &TW&     0.723&     0.470\\
          &LMC&     -3.424$\pm$0.018     &     0.104&   1049&                        &TW&       ---&       ---\\
 $W_{V,I}$&SMC&     -3.626$\pm$0.020     &     0.153&   1526&     0.160$\pm$0.023     &TW&     6.193&     0.000\\
          &LMC&     -3.466$\pm$0.011     &     0.069&   1075&                        &B15&       ---&       ---\\
\hline
\end{tabular}
\end{center}
{\footnotesize \textbf{Notes}: Source: TW - this work; M15 - \citet{macri15}; B15 - \citet{cpapir3}. $^a\Delta$ 
is the difference in the slopes of LMC and SMC PL relations. The error ($\sigma_{\Delta}^b$) is 
obtained using the quadrature sum of the errors in two slopes.}
\end{minipage}
\end{table*}

We cross-matched OGLE-III LMC FO mode Cepheids with publicly released SAGE data using a search radius of $2''$ and 
obtained IRAC band photometry for epoch 1 and 2. The number of matched sources and 
the corresponding mean separations and standard deviations are summarized in Table~\ref{table:irac_delta1} 
for the SAGE data. We estimated the error weighted mean in case the magnitudes were available for more than one 
epoch of observation for a particular Cepheid. Similar to SMC FO mode Cepheids, the magnitudes in $5.8\mu \mathrm{m}$ and 
$8.0\mu \mathrm{m}$-bands for LMC FO mode Cepheids are mostly affected by the incompleteness bias at short period ends and 
therefore, we do not consider these bands in deriving PL relations. We only make use of magnitudes in 
$3.6\mu \mathrm{m}$ and $4.5\mu \mathrm{m}$-bands and correct them for extinction using the Haschke maps \citep{hasch11}
as discussed previously. We restrict our sample for $P>0.6$~days because there are very few stars below
this period and apply recursive $2.5\sigma$ clipping before fitting PL relations.

Fig.~\ref{fig:lmc_mir} displays the PL relations for LMC FO mode Cepheids at optical and mid-infrared 
wavelengths and the optical Wesenheit relation while the results are presented in Table~\ref{table:lmc_mir}. 
We note that the dispersion in these PL relations is significantly smaller than those for the SMC FO mode
PL relations. As suggested previously, the increased dispersion for the SMC FO mode PL relations may be 
attributed to the high line-of-sight depth of the galaxy \citep{subra15}. We note that \citet{cpapir3} found 
a significant break in optical band relations at $\log P = 0.4$ for LMC FO mode Cepheids but no such break is 
observed in near-infrared PL relations. We also test for possible non-linearity in PL relations derived in
this study for FO mode Cepheids at 2.5 days. The results of the F-test, random walk and the testimator are
presented in Table~\ref{table:fo_lmc_frw} and \ref{table:fo_lmc_tm}. The optical band PL relations 
provide significant change in slope around 2.5 days, similar to SMC FO mode Cepheids. Interestingly,
the mid-infrared PL relations also provide evidence of a significant break at $\log P =0.4$. The F and random
walk test suggest the probabilities of acceptance of null hypothesis i.e. a linear relation, are approximately
zero. Therefore, a significant change is observed in the slope of the PL relations at 2.5 days for LMC FO mode
Cepheids at $3.6$ \& $4.5\mu \mathrm{m}$ wavelengths.

We provide a detailed comparison of the slopes for FO mode Cepheid PL relations between the LMC and SMC
in the period range $0.6<P<6.3$~days and the results are presented in Table~\ref{table:comp_slopes}.
The t-test suggests that the optical $V$-band and the Wesenheit function are not consistent
in the two clouds, while the $I$-band PL relations exhibit similar slopes with a marginal difference. Most of these 
relations in the infrared are consistent, given the uncertainties in the slopes, except in the case of the $J$ and 
$3.6\mu \mathrm{m}$ band PL relations. The difference in the slopes for the two clouds is least in $I$-band 
and $4.5\mu\mathrm{m}$-band PL relations.

\subsection{Distance modulus between the SMC and LMC using FO mode Cepheids}

We determine the relative distance modulus between the SMC and LMC based on multiband P-L relations for FO
mode Cepheids. We fit an equation to each band PL relations in the Magellanic Clouds in the following form :

\begin{gather}
\label{eq:linearpl}
m = (\mu_{SMC} - \mu_{LMC}) + b_{LMC} + a_S\log P_S + a_L\log P_L,
\end{gather}

\noindent where, $m$ is the magnitude for Cepheids in the Magellanic Clouds in a particular band. The first 
term, $(\mu_{SMC} - \mu_{LMC})$, provides the relative distance modulus between the Magellanic Clouds. The coefficients, $a$ and 
$b$ represent the slope and intercept, respectively and the subscript $S$ and $L$ represent the short (P$<$2.5 days) 
and long (P$>$2.5 days) period Cepheids. We use two slope linear regression in our analysis as optical band PL 
relations provide evidence of statistically significant non-linearities at 2.5 days.

\begin{table}
\begin{minipage}{1.0\hsize}
\begin{center}
\caption{ Distance modulus between LMC and SMC estimated using FO mode Cepheids. \label{table:diff_mu}}
\begin{tabular}{|l|c|c|c|}
\hline
\hline
Band	&		$a_S$& 		$a_L$&		$\mu_{SMC} - \mu_{LMC}$\\
\hline
\hline
                 $V$&     -3.301$\pm$0.030     &     -2.782$\pm$0.093     &      0.422$\pm$0.013     \\
                 $I$&     -3.418$\pm$0.023     &     -2.962$\pm$0.073     &      0.462$\pm$0.010     \\
                 $J$&     -3.239$\pm$0.059     &     -2.990$\pm$0.097     &      0.448$\pm$0.018     \\
                 $H$&     -3.204$\pm$0.050     &     -3.121$\pm$0.090     &      0.532$\pm$0.015     \\
             $K_{s}$&     -3.303$\pm$0.062     &     -3.099$\pm$0.077     &      0.495$\pm$0.014     \\
         $3.6\mu \mathrm{m}$&     -3.534$\pm$0.022     &     -3.177$\pm$0.054     &      0.513$\pm$0.009     \\
         $4.5\mu \mathrm{m}$&     -3.494$\pm$0.026     &     -3.178$\pm$0.057     &      0.497$\pm$0.010     \\
           $W_{V,I}$&     -3.621$\pm$0.016     &     -3.266$\pm$0.050     &      0.519$\pm$0.007     \\
\hline
		    &			       &                          &      $\Delta\mu=0.49\pm0.02$\\ 
\hline
\end{tabular}
\end{center}
{\footnotesize {{\bf Note :} The last column represents the relative distance modulus. $a$ is the slope and subscripts 
$S$ and $L$ represent short ($P<2.5$~days) and long ($P>2.5$~days) period range Cepheids.}}
\end{minipage}
\end{table}

We provide the slopes and relative distance moduli between SMC and LMC in Table~\ref{table:diff_mu}. We 
note that the difference in the relative distance modulus varies significantly from optical to infrared 
bands. We estimate the mean and standard deviation of all relative distances and adopt an average relative distance 
modulus of $\Delta \mu = 0.49\pm0.02$. This value is consistent with those based on FU mode Cepheids ($0.48\pm0.02$) 
derived by \citet{ngeow15}. We also note that \citet{delmc} and \citet{desmc} recommended a distance modulus 
of $18.49\pm0.09$~mag and $18.96\pm0.02$~mag for LMC and SMC, respectively. The relative distance modulus 
between the two clouds based on these recommended distances is $0.47\pm0.09$~mag. Therefore, our estimated 
relative distance modulus is in good agreement with the difference in recommended distance moduli of the two 
Clouds.

\section{Conclusions}
\label{sec:discuss}

In this work, we derive new multi-band PL relations for first-overtone mode Cepheids in the SMC. In addition to a large
compilation of OGLE-III Cepheids, we use their counterparts in 2MASS and SAGE catalogs to derive multi-band
mean magnitudes for Cepheids in the SMC. The extinction corrections are done using the Haschke maps \citep{hasch11}.
We also extend the work of \citet{ngeow15} to include short period SMC fundamental-mode Cepheids in our analysis. 
We use robust statistical tests such as, F-test, random walk test and the testimator,
to determine the significance of possible non-linearities at various periods in PL 
relations for fundamental and first-overtone mode Cepheids in the SMC. We also derive new optical and 
mid-infrared band PL relations for first-overtone mode Cepheids in the LMC. The first-overtone mode PL 
relation in the Magellanic Clouds are compared and we find that most of these relations are not consistent 
in the two clouds. We summarize the main results from our study as follows.

\begin{itemize}

\item{The multi-band PL relations for FO mode Cepheids in the Magellanic Clouds, derived in this study, are found to be consistent with
previous studies in the literature. The updated infrared band PL relations for FO mode Cepheids are not studied in 
detail previously, and will be highly relevant in the upcoming era of the {\it James Webb Space Telescope}.}

\item{We find significant evidence of a break in PL relations for both fundamental and first-overtone 
mode Cepheids in the SMC at 2.5 days at optical bands.
The fundamental-mode SMC Cepheids also exhibit this break in infrared 
band PL relations. We do not find any evidence of non-linearity at 1 days as observed by \citet{subra15} for first-overtone mode SMC Cepheids.}

\item{Our analysis suggests that these breaks are related to changes in the progression of light curve parameters 
with period for Classical Cepheids in the SMC. This is an important result in a sense that modelling these non-linearities 
in PL relations, i.e. physical parameters, together with changes in the Fourier parameters will be used to constrain 
theoretical pulsation codes and determine mass-luminosity relations obeyed by Cepheids in the Magellanic Clouds. }

\item{We also compare the multi-band PL relations in the Magellanic clouds for first-overtone mode Cepheids. The slope
of the PL relations in the two clouds are not consistent at optical bands, presumably due to significant non-linearities.
We estimate the relative distance modulus of $\Delta\mu=0.49\pm0.02$~mag, between the two clouds using multi-band PL 
relations for first-overtone mode Cepheids in the SMC and LMC.}

\item{The various non-linearities and the inconsistency in the slopes between the two clouds may be related
to the metallicity differences thus leading to different Cepheid mass-luminosity relations. \citet{bhardwaj14}
 have provided a possible explanation for breaks in Period-Color relations at a period of 10 days based on the theory of  
hydrogen ionization front-stellar photosphere interaction as a function of pulsation phase. 
Further study will be required to discern the cause of nonlinearities at shorter periods. In future, with
more infrared data coming from The VISTA near-infrared $YJK_s$ survey of the Magellanic System (VMC) survey
\citep{cioni2011, ripepi2012, ripepi2016}, these relations will be studied as a function of pulsation phase 
to understand the metallicity dependence on the universality of Cepheid PL relations.}

\end{itemize}

\section*{Acknowledgments}
\label{sec:ackno}

We thank the anonymous referee for his/her valuable comments which improved the content and quality of the manuscript.
AB is thankful to the Council of Scientific and Industrial Research, New Delhi, for the Senior Research 
Fellowship grant 09/045(1296)/2013-EMR-I.  This work is supported by the grant for the Joint Center for 
Analysis of Variable Star Data provided by Indo-U.S. Science and Technology Forum. 
CCN thanks the funding from Ministry of Science and Technology (Taiwan) under the contract NSC101-2112-M-008-017-MY3
and NSC104-2112-M-008-012-MY3. HPS thanks University of Delhi for a R\&D grant. 
This work also makes use of data products from the 2MASS survey, which is a joint project of the University of 
Massachusetts and the Infrared Processing and Analysis Center/California Institute of Technology, 
funded by the National Aeronautics and Space Administration and the National Science Foundation.
In addition, this study also makes use of NASA's Astrophysics Data System, 
the VizieR catalogues. 


\bibliographystyle{mn2e}
\bibliography{SMC_FO}

\begin{thebibliography}{43}
\expandafter\ifx\csname natexlab\endcsname\relax\def\natexlab#1{#1}\fi


\bibitem[{{Bauer} {et~al}\mbox{.}(1999){Bauer}, {Afonso}, {Albert}, {Alard},
  {Andersen}, {Ansari}, {Aubourg}, {Bareyre}, {Beaulieu}, {Bouquet}, {Char},
  {Charlot}, {Couchot}, {Coutures}, {Derue}, {Ferlet}, {Gaucherel},
  {Glicenstein}, {Goldman}, {Gould}, {Graff}, {Gros}, {Haissinski}, {Hamilton},
  {Hardin}, {de Kat}, {Kim}, {Lasserre}, {Lesquoy}, {Loup}, {Magneville},
  {Mansoux}, {Marquette}, {Maurice}, {Milsztajn}, {Moniez},
  {Palanque-Delabrouille}, {Perdereau}, {Pr{\'e}vot}, {Renault}, {Regnault},
  {Rich}, {Spiro}, {Vidal-Madjar}, {Vigroux}, \& {Zylberajch}}]{bauer99}
{Bauer} F. {et~al.}, 1999, AAP, 348, 175

\bibitem[{{Bhardwaj} {et~al}\mbox{.}(2014){Bhardwaj}, {Kanbur}, {Singh}, \&
  {Ngeow}}]{bhardwaj14}
{Bhardwaj} A., {Kanbur} S.~M., {Singh} H.~P., {Ngeow} C.-C., 2014, MNRAS, 445,
  2655

\bibitem[{{Bhardwaj} {et~al}\mbox{.}(2015{\natexlab{a}}){Bhardwaj}, {Kanbur},
  {Macri}, {Singh}, {Ngeow}, {Wagner-Kaiser}, \& {Sarajedini}}]{cpapir2}
{Bhardwaj} A., {Kanbur} S.~M., {Macri} L.~M., {Singh} H.~P., {Ngeow} C.-C.,
  {Wagner-Kaiser} R., {Sarajedini} A., 2015{\natexlab{a}}, Accepted in AJ,
  ArXiv e-prints, 1510.03682

\bibitem[{{Bhardwaj} {et~al}\mbox{.}(2015{\natexlab{b}}){Bhardwaj}, {Kanbur},
  {Singh}, {Macri}, \& {Ngeow}}]{bhardwaj15}
{Bhardwaj} A., {Kanbur} S.~M., {Singh} H.~P., {Macri} L.~M., {Ngeow} C.-C.,
  2015{\natexlab{b}}, MNRAS, 447, 3342

\bibitem[{{Bhardwaj} {et~al}\mbox{.}(2016){Bhardwaj}, {Kanbur}, {Macri},
  {Singh}, {Ngeow}, \& {Ishida}}]{cpapir3}
{Bhardwaj} A., {Kanbur} S.~M., {Macri} L.~M., {Singh} H.~P., {Ngeow} C.-C.,
  {Ishida} E.~E.~O., 2016, MNRAS, 457, 1644

\bibitem[{{Bono} {et~al}\mbox{.}(2002){Bono}, {Groenewegen}, {Marconi}, \&
  {Caputo}}]{bono02}
{Bono} G., {Groenewegen} M.~A.~T., {Marconi} M., {Caputo} F., 2002, ApJL, 574,
  L33

\bibitem[{{Caldwell} \& {Laney}(1991)}]{cald91}
{Caldwell} J.~A.~R., {Laney} C.~D., 1991, in IAU Symposium, Vol. 148, The
  Magellanic Clouds, {Haynes} R., {Milne} D., eds., p. 249

\bibitem[{{Cardelli}, {Clayton} \& {Mathis}(1989){Cardelli}, {Clayton}, \&
  {Mathis}}]{card89}
{Cardelli} J.~A., {Clayton} G.~C., {Mathis} J.~S., 1989, APJ, 345, 245

\bibitem[{{Cioni} {et~al}\mbox{.}(2011){Cioni}, {Clementini}, {Girardi},
  {Guandalini}, {Gullieuszik}, {Miszalski}, {Moretti}, {Ripepi}, {Rubele},
  {Bagheri}, {Bekki}, {Cross}, {de Blok}, {de Grijs}, {Emerson}, {Evans},
  {Gibson}, {Gonzales-Solares}, {Groenewegen}, {Irwin}, {Ivanov}, {Lewis},
  {Marconi}, {Marquette}, {Mastropietro}, {Moore}, {Napiwotzki}, {Naylor},
  {Oliveira}, {Read}, {Sutorius}, {van Loon}, {Wilkinson}, \&
  {Wood}}]{cioni2011}
{Cioni} M.-R.~L. {et~al.}, 2011, A\&A, 527, A116

\bibitem[{{Cutri} {et~al}\mbox{.}(2003){Cutri}, {Skrutskie}, {van Dyk},
  {Beichman}, {Carpenter}, {Chester}, {Cambresy}, {Evans}, {Fowler}, {Gizis},
  {Howard}, {Huchra}, {Jarrett}, {Kopan}, {Kirkpatrick}, {Light}, {Marsh},
  {McCallon}, {Schneider}, {Stiening}, {Sykes}, {Weinberg}, {Wheaton},
  {Wheelock}, \& {Zacarias}}]{cutri03}
{Cutri} R.~M. {et~al.}, 2003, VizieR Online Data Catalog, 2246, 0

\bibitem[{{de Grijs}, {Wicker} \& {Bono}(2014){de Grijs}, {Wicker}, \&
  {Bono}}]{delmc}
{de Grijs} R., {Wicker} J.~E., {Bono} G., 2014, AJ, 147, 122

\bibitem[{{de Grijs} \& {Bono}(2015)}]{desmc}
{de Grijs} R., {Bono} G., 2015, AJ, 149, 179

\bibitem[{{Deb} \& {Singh}(2009)}]{deb09}
{Deb} S., {Singh} H.~P., 2009, A\&A, 507, 1729

\bibitem[{{Groenewegen}(2000)}]{gmat00}
{Groenewegen} M.~A.~T., 2000, A\&A, 363, 901

\bibitem[{{Haschke}, {Grebel} \& {Duffau}(2011){Haschke}, {Grebel}, \&
  {Duffau}}]{hasch11}
{Haschke} R., {Grebel} E.~K., {Duffau} S., 2011, AJ, 141, 158

\bibitem[{{Inno} {et~al}\mbox{.}(2013){Inno}, {Matsunaga}, {Bono}, {Caputo},
  {Buonanno}, {Genovali}, {Laney}, {Marconi}, {Piersimoni}, {Primas}, \&
  {Romaniello}}]{inno13}
{Inno} L. {et~al.}, 2013, ApJ, 764, 84

\bibitem[{{Inno} {et~al}\mbox{.}(2015){Inno}, {Matsunaga}, {Romaniello},
  {Bono}, {Monson}, {Ferraro}, {Iannicola}, {Persson}, {Buonanno}, {Freedman},
  {Gieren}, {Groenewegen}, {Ita}, {Laney}, {Lemasle}, {Madore}, {Nagayama},
  {Nakada}, {Nonino}, {Pietrzy{\'n}ski}, {Primas}, {Scowcroft},
  {Soszy{\'n}ski}, {Tanab{\'e}}, \& {Udalski}}]{inno15}
{Inno} L. {et~al.}, 2015, A\&A, 576, A30

\bibitem[{{{\it Planck} collaboration, Ade} {et~al}\mbox{.}(2014){{\it Planck}
  collaboration, Ade}, {Aghanim}, {Armitage-Caplan}, {Arnaud}, {Ashdown},
  {Atrio-Barandela}, {Aumont}, {Baccigalupi}, {Banday}, \& et~al.}]{planck14}
{{\it Planck} collaboration, Ade} P.~A.~R. {et~al.}, 2014, A\&A, 571, A16

\bibitem[{{Laney} \& {Stobie}(1986)}]{laney86}
{Laney} C.~D., {Stobie} R.~S., 1986, MNRAS, 222, 449

\bibitem[{{Laney} \& {Stobie}(1994)}]{laney94}
{Laney} C.~D., {Stobie} R.~S., 1994, MNRAS, 266, 441

\bibitem[{{Leavitt} \& {Pickering}(1912)}]{leavitt12}
{Leavitt} H.~S., {Pickering} E.~C., 1912, Harvard College Observatory Circular,
  173, 1

\bibitem[{{Macri} {et~al}\mbox{.}(2015){Macri}, {Ngeow}, {Kanbur}, {Mahzooni},
  \& {Smitka}}]{macri15}
{Macri} L.~M., {Ngeow} C.-C., {Kanbur} S.~M., {Mahzooni} S., {Smitka} M.~T.,
  2015, AJ, 149, 117

\bibitem[{{Matsunaga}, {Feast} \& {Soszy{\'n}ski}(2011){Matsunaga}, {Feast}, \&
  {Soszy{\'n}ski}}]{matsunaga11}
{Matsunaga} N., {Feast} M.~W., {Soszy{\'n}ski} I., 2011, MNRAS, 413, 223

\bibitem[{{Ngeow} \& {Kanbur}(2010)}]{ngeow10}
{Ngeow} C.-C., {Kanbur} S.~M., 2010, ApJ, 720, 626

\bibitem[{{Ngeow}, {Citro} \& {Kanbur}(2012){Ngeow}, {Citro}, \&
  {Kanbur}}]{ngeow12}
{Ngeow} C.-C., {Citro} D.~M., {Kanbur} S.~M., 2012, MNRAS, 420, 585

\bibitem[{{Ngeow} {et~al}\mbox{.}(2015){Ngeow}, {Kanbur}, {Bhardwaj}, \&
  {Singh}}]{ngeow15}
{Ngeow} C.-C., {Kanbur} S.~M., {Bhardwaj} A., {Singh} H.~P., 2015, ApJ, 808, 67

\bibitem[{{Riess} {et~al}\mbox{.}(2009){Riess}, {Macri}, {Casertano}, {Sosey},
  {Lampeitl}, {Ferguson}, {Filippenko}, {Jha}, {Li}, {Chornock}, \&
  {Sarkar}}]{riess09}
{Riess} A.~G. {et~al.}, 2009, ApJ, 699, 539

\bibitem[{{Riess} {et~al}\mbox{.}(2011){Riess}, {Macri}, {Casertano},
  {Lampeitl}, {Ferguson}, {Filippenko}, {Jha}, {Li}, \& {Chornock}}]{riess11}
{Riess} A.~G. {et~al.}, 2011, ApJ, 730, 119

\bibitem[{{Ripepi} {et~al}\mbox{.}(2012){Ripepi}, {Moretti}, {Marconi},
  {Clementini}, {Cioni}, {Marquette}, {Girardi}, {Rubele}, {Groenewegen}, {de
  Grijs}, {Gibson}, {Oliveira}, {van Loon}, \& {Emerson}}]{ripepi2012}
{Ripepi} V. {et~al.}, 2012, MNRAS, 424, 1807

\bibitem[{{Ripepi} {et~al}\mbox{.}(2016){Ripepi}, {Marconi}, {Moretti},
  {Clementini}, {Cioni}, {de Grijs}, {Emerson}, {Groenewegen}, {Ivanov}, \&
  {Piatti}}]{ripepi2016}
{Ripepi} V. {et~al.}, 2016, arXiv:1602.09005

\bibitem[{{Sandage}(1988)}]{sandage88}
{Sandage} A., 1988, PASP, 100, 935

\bibitem[{{Sandage}, {Tammann} \& {Reindl}(2009){Sandage}, {Tammann}, \&
  {Reindl}}]{tammann09}
{Sandage} A., {Tammann} G.~A., {Reindl} B., 2009, A\&A, 493, 471

\bibitem[{{Sharpee} {et~al}\mbox{.}(2002){Sharpee}, {Stark}, {Pritzl}, {Smith},
  {Silbermann}, {Wilhelm}, \& {Walker}}]{sharp02}
{Sharpee} B., {Stark} M., {Pritzl} B., {Smith} H., {Silbermann} N., {Wilhelm}
  R., {Walker} A., 2002, AJ, 123, 3216

\bibitem[{{Soszy{\'n}ski}, {Gieren} \& {Pietrzy{\'n}ski}(2005){Soszy{\'n}ski},
  {Gieren}, \& {Pietrzy{\'n}ski}}]{sosz05}
{Soszy{\'n}ski} I., {Gieren} W., {Pietrzy{\'n}ski} G., 2005, PASP, 117, 823

\bibitem[{{Soszy{\~n}ski} {et~al}\mbox{.}(2010){Soszy{\~n}ski}, {Poleski},
  {Udalski}, {Szyma{\~n}ski}, {Kubiak}, {Pietrzy{\~n}ski}, {Wyrzykowski},
  {Szewczyk}, \& {Ulaczyk}}]{oglesmcceph}
{Soszy{\~n}ski} I. {et~al.}, 2010, Acta Astron., 60, 17

\bibitem[{{Storm} {et~al}\mbox{.}(2004){Storm}, {Carney}, {Gieren},
  {Fouqu{\'e}}, {Latham}, \& {Fry}}]{storm04}
{Storm} J., {Carney} B.~W., {Gieren} W.~P., {Fouqu{\'e}} P., {Latham} D.~W.,
  {Fry} A.~M., 2004, A\&A, 415, 531

\bibitem[{{Subramanian} \& {Subramaniam}(2015)}]{subra15}
{Subramanian} S., {Subramaniam} A., 2015, A\&A, 573, A135

\bibitem[{{Tammann}, {Sandage} \& {Reindl}(2003){Tammann}, {Sandage}, \&
  {Reindl}}]{tammann03}
{Tammann} G.~A., {Sandage} A., {Reindl} B., 2003, A\&A, 404, 423

\bibitem[{{Tammann}, {Reindl} \& {Sandage}(2011){Tammann}, {Reindl}, \&
  {Sandage}}]{tammann11}
{Tammann} G.~A., {Reindl} B., {Sandage} A., 2011, A\&A, 531, A134

\bibitem[{{Udalski} {et~al}\mbox{.}(1999{\natexlab{a}}){Udalski}, {Soszynski},
  {Szymanski}, {Kubiak}, {Pietrzynski}, {Wozniak}, \& {Zebrun}}]{udal99}
{Udalski} A., {Soszynski} I., {Szymanski} M., {Kubiak} M., {Pietrzynski} G.,
  {Wozniak} P., {Zebrun} K., 1999{\natexlab{a}}, Acta Astron., 49, 437

\bibitem[{{Udalski} {et~al}\mbox{.}(1999{\natexlab{b}}){Udalski}, {Szymanski},
  {Kubiak}, {Pietrzynski}, {Soszynski}, {Wozniak}, \& {Zebrun}}]{udal99a}
{Udalski} A., {Szymanski} M., {Kubiak} M., {Pietrzynski} G., {Soszynski} I.,
  {Wozniak} P., {Zebrun} K., 1999{\natexlab{b}}, Acta Astron., 49, 201

\bibitem[{{Welch} \& {Madore}(1984)}]{welch84}
{Welch} D.~L., {Madore} B.~F., 1984, in IAU Symposium, Vol. 108, Structure and
  Evolution of the Magellanic Clouds, {van den Bergh} S., {de Boer} K.~S.~D.,
  eds., p. 221

\bibitem[{{Welch} {et~al}\mbox{.}(1987){Welch}, {McLaren}, {Madore}, \&
  {McAlary}}]{welch87}
{Welch} D.~L., {McLaren} R.~A., {Madore} B.~F., {McAlary} C.~W., 1987, ApJ,
  321, 162

\end{thebibliography}

\end{document}